\documentclass[acmsmall,natbib]{acmart}
\AtBeginDocument{%
  }

\setcopyright{acmlicensed}
\copyrightyear{xxxx}
\acmYear{2025}
\acmDOI{XXXXXXX.XXXXXXX}

\acmSubmissionID{XXX-XXX-XXX}

\usepackage[capitalise]{cleveref}
\usepackage{natbib}
\begin{document}

\title{Algorithmic UDAP}

\author{Talia Gillis}

\affiliation{%
  \institution{Columbia University}
  \city{New York}
  \state{New York}
  \country{USA}
}
\email{tbg2117@columbia.edu}

\author{Riley Stacy}
\affiliation{%
  \institution{Barnard College}
  \city{New York}
  \state{New York}
  \country{USA}
}
\email{rs4142@barnard.edu}

\author{Sam Brumer}
\affiliation{%
  \institution{Columbia University}
  \city{New York}
  \state{New York}
  \country{USA}
}
\email{sdb2184@columbia.edu}

\author{Emily Black}
\affiliation{%
  \institution{NYU}
  \city{New York}
  \state{New York}
  \country{USA}
}
\email{emilyblack@nyu.edu}


\begin{abstract}
  This paper compares two legal frameworks---disparate impact (DI) and unfair, deceptive, or abusive acts or practices (UDAP)---as tools for evaluating algorithmic discrimination, focusing on the example of fair lending. While DI has traditionally served as the foundation of fair lending law, recent regulatory efforts have invoked UDAP, a doctrine rooted in consumer protection, as an alternative means to address algorithmic discrimination harms. We formalize and operationalize both doctrines in a simulated lending setting to assess how they evaluate algorithmic disparities. While some regulatory interpretations treat UDAP as operating similarly to DI, we argue it is an independent and analytically distinct framework. In particular, UDAP’s ``unfairness'' prong introduces elements such as avoidability of harm and proportionality balancing, while its ``deceptive'' and ``abusive'' standards may capture forms of algorithmic harm that elude DI analysis. At the same time, translating UDAP into algorithmic settings exposes unresolved ambiguities, underscoring the need for further regulatory guidance if it is to serve as a workable standard.  
\end{abstract}


\received{30 September 2025}
\received[accepted]{5 December 2025}

\maketitle

\section{Introduction}
For decades, the disparate impact (DI) doctrine has served as the cornerstone of fair lending law, providing a framework for identifying and remedying discriminatory practices that disproportionately affect protected groups, even in the absence of intentional discrimination. Under statutes like the Fair Housing Act (FHA) and the Equal Credit Opportunity Act (ECOA), the DI framework follows a well-established three-part burden-shifting test: plaintiffs must first demonstrate that a facially neutral practice produces disparate outcomes for protected groups; defendants can then assert a business justification for the practice; and finally, plaintiffs may show that the same business interest could be served by a less discriminatory alternative \citep{black2023less,gillis2024operationalizing,gillis2025price}.


Recent regulatory developments have reshaped the legal landscape for addressing algorithmic discrimination. Under the Biden administration, the Consumer Financial Protection Bureau (CFPB) and Federal Trade Commission (FTC) advanced the idea that the prohibition on unfair, deceptive, or abusive acts or practices (UDAP),\footnote{As will be explained below, Section 5 of the Federal Trade Commission Act \citep{FTCA1914} prohibits unfair and deceptive acts or practices and is typically referred to as ``UDAP’’. Section 1031 of the Dodd-Frank Act \citep{CFPA2010}, prohibits unfair, deceptive, and/or abusive acts or practices and is typically referred to as ``UDAAP''.  Because this article focuses primarily on the ``unfairness’’ prong, for simplicity, we use the term ``UDAP'' to refer collectively to the FTC Act and Dodd-Frank provisions, even though Dodd-Frank's provision includes an ``abusive'' prong. When discussing Dodd-frank specifically we may refer to the provision as ``UDAAP’’.} originally rooted in consumer protection law, could serve as an additional tool for addressing discrimination, including algorithmic discrimination in lending. By contrast, the current Trump administration has signaled growing skepticism toward disparate impact liability\footnote{See recent Executive Order, \citet{WhiteHouse2025EO14281}. In the first Trump administration, there were also attempts to pull back on DI enforcement. See then-Acting Director Mick Mulvaney's statement that the CFPB will reexamine its guidance on disparate impact liability under the ECOA \citep{cfpb_statement_sj_res_57_2018}.} and toward expansive uses of UDAP in the anti-discrimination context. Together, these shifts point in divergent directions, creating uncertainty about which legal framework will guide the evaluation of algorithmic lending practices and underscoring the need to clarify how DI and UDAP differ in structure, scope, and normative foundation.

These developments raise fundamental questions about the relationship between UDAP and DI. While some regulatory interpretations treat UDAP as operating similarly to DI, we argue that the underlying structure and normative foundations of these frameworks differ. UDAP's ``unfairness'' prong requires showing that a practice causes substantial injury that is not reasonably avoidable and is not outweighed by countervailing benefits---elements that differ from traditional DI analysis. Meanwhile, UDAP's ``deceptive'' and ``abusive'' standards may capture forms of algorithmic harm that fall outside the scope of DI entirely.

Although UDAP has been identified as a possible tool against algorithmic discrimination~\citep{hirsch2014s,SelbstBarocas2023}, its practical implications remain largely unexplored, particularly in contexts where machine learning’s complexity intersects with shifting legal standards. As lenders increasingly deploy sophisticated algorithms for credit decisions, understanding how different legal frameworks would evaluate the same algorithmic system becomes crucial for both compliance and policy development. Yet the technical operationalization of these legal standards---translating abstract legal concepts into concrete, measurable criteria---has received limited attention in existing scholarship.

These doctrinal differences are particularly important in the context of algorithmic lending. As financial institutions increasingly deploy complex machine learning models for credit decisions, the question is how legal standards can be translated into workable criteria for evaluating these systems. Yet the technical operationalization of doctrines such as disparate impact and UDAP---translating abstract legal concepts into concrete, measurable criteria---has received limited attention in existing scholarship.

This paper addresses this gap by providing the first \textit{technical} comparison of DI and UDAP as frameworks for evaluating algorithmic discrimination in lending. We make several key contributions. First, we formalize both legal doctrines in technical terms, creating operational criteria that can be applied to algorithmic systems. Second, we demonstrate through simulation how these frameworks would evaluate the same lending algorithm, revealing where they converge and diverge in their assessment of algorithmic practices. Third, we identify critical ambiguities and open questions that must be resolved through regulatory guidance to make UDAP a workable standard for algorithmic governance.

Our analysis reveals that while UDAP and DI may sometimes reach similar conclusions about particular algorithmic practices, they reflect fundamentally different approaches to evaluating potential harms. UDAP's incorporation of proportionality analysis through its requirement that harms not be outweighed by countervailing benefits introduces a form of cost-benefit balancing that is largely absent from traditional DI doctrine. This difference has significant implications for how algorithmic systems are evaluated and what kinds of justifications are deemed acceptable for practices that produce disparate outcomes.

Moreover, our technical implementation highlights numerous areas where existing UDAP guidance provides insufficient clarity for algorithmic applications. Questions such as what constitutes ``substantial injury'' in the context of algorithmic lending decisions, how to assess whether discriminatory harms are ``reasonably avoidable,'' and how to weigh business benefits against discriminatory costs under the ``unfairness'' prong of UDAP remain largely unresolved. These ambiguities create significant uncertainty for both regulated entities seeking to ensure compliance and enforcers attempting to apply the doctrine consistently.


Our focus on better understanding the UDAP framework as a tool for addressing discriminatory conduct is particularly timely. Decades of regulatory enforcement and case law have recognized DI under the Equal Credit Opportunity Act (ECOA) and the Fair Housing Act (FHA), but its legal foundation has recently come under sustained challenge, raising doubts about its continued robustness. In this environment, exploring UDAP not merely as a substitute but as a complementary doctrinal foundation helps identify paths for sustained fair lending enforcement.
 
These developments suggest that fair lending protections may benefit from being grounded in multiple legal bases to ensure continued enforcement against discriminatory practices. 
In addition, many states have their own statutes prohibiting unfair and deceptive conduct, which grant them enforcement discretion beyond federal oversight. 
As a result, much of the analysis presented here may also inform the interpretation and application of state UDAP\footnote{State laws typically do not include a prohibition on ``abusive'' conduct.} laws \citep{Sovern2024}. In this context, UDAP represents a potential complementary tool for addressing algorithmic discrimination. However, realizing this potential requires careful attention to how the doctrine should be operationalized and where regulatory clarification is most urgently needed.

\paragraph{Related Literature}

This paper contributes to scholarship on how discrimination law applies to algorithmic decision-making and what forms of regulatory oversight it requires. Much of this work has focused on the DI doctrine, especially in lending \citep{gillis2021input}, employment \citep{kim2016data,kim2022race}, and housing \citep{foggo2020algorithms}. While the technical fairness literature proposes numerous ways to measure disparities \citep{verma2018fairness,caro2025differential}, only a smaller body of scholarship connects these measures to the doctrinal details of DI. Within this line of work, scholars have both critiqued existing doctrine and proposed clarifications needed to address algorithmic harms \citep{black2023less,black2024d}. Our paper extends this conversation by comparing DI to UDAAP, showing how each doctrine would evaluate algorithmic disparities.

Much scholarship evaluates the consumer protection powers of UDAPs generally, surveying real and proposed laws for their strengths and weaknesses \citep{nclc_maps_how_well_states_protect_consumers_2018,nclc_unfair_deceptive_acts_practices_2021,pridgen2015wrecking,faust2023regulating,cooper2016state}. Other works focus specifically on state UDAPs, examining their legislative history and development as part of the states'  and nation’s political landscape \citep{gilles2023private,cooper2016state,orrick_consumer_financial_services_2010,pridgen2016dynamic,pridgen2015wrecking,cox2018strategies}. Finally, recent scholarship has evaluated the relationship between UDAP unfairness regulation and discrimination, both generally \citep{Sovern2024,elengold2019consumer,HayesSchellenberg2021} and in the context of artificial intelligence \citep{hirsch2014s,SelbstBarocas2023}.

\paragraph{Outline} This paper proceeds in three parts. \Cref{sec:legal} provides an overview of both DI and UDAP doctrines, examining their legal foundations, analytical frameworks, and existing regulatory guidance. \Cref{sec:tehcnical} presents our technical implementation of DI and the ``unfairness'' prong of UDAP, using simulation to show how they would evaluate a hypothetical algorithmic lending system and to highlight points of convergence and divergence. We also identify key ambiguities and gaps in regulatory guidance. Finally,  \Cref{sec:future_guidance} considers broader implications of our analysis, including the emerging importance of state UDAPs and the potential for UDAP's ``deceptive'' and ``abusive'' prongs to address algorithmic harms, including discrimination. 

Through this analysis, we aim to contribute to both legal scholarship and policy development by providing a foundation for understanding how different legal frameworks approach algorithmic discrimination. As the use of algorithmic decision-making continues to expand across the financial services sector, developing clear and workable legal standards for evaluating these systems becomes increasingly critical for ensuring both innovation and fairness in consumer lending markets.

\section{Legal Foundations: Disparate Impact and UDAAP as Anti-Discrimination Tools}
\label{sec:legal}
\subsection{Legal Background}

\subsubsection{Disparate Impact.} The two federal statutes that form the core prohibition on discrimination in credit are the Fair Housing Act (FHA) \citep{ECOA1974} and the Equal Credit Opportunity Act (ECOA) \citep{ECOA1974}.  FHA, also known as Title VIII of the Civil Rights Act of 1968, protects buyers and renters from discrimination by sellers or landlords. It covers a range of housing-related conduct and prohibits discrimination in setting housing-related credit terms based on race, color, religion, sex, disability, familial status, and national origin.  ECOA prohibits discrimination in all credit transactions, beyond those in the context of housing. Both ECOA and FHA incorporate the doctrines of ``disparate treatment'' and ``disparate impact.''\footnote{While the texts of ECOA and FHA do not explicitly recognize the two discrimination doctrines, the disparate impact doctrine has long been recognized in credit pricing cases by courts and agencies alike. The Supreme Court affirmed that disparate impact claims could be made under FHA in \textit{Texas Dep’t of Hous. and Cmty. Affs. v. Inclusive Communities Project, Inc.}, 576 U.S. 519 (2015) \citep{InclusiveCommunities2015}, confirming the position of eleven appellate courts and various federal agencies, including the Department of Housing and Urban Development (HUD), the agency primarily responsible for enforcing FHA \citep{schwemm2015fair}. There is no equivalent Supreme Court case with respect to ECOA, but the Consumer Financial Protection Bureau (CFPB), the agency primarily responsible for enforcing the ECOA, and lower courts have found that the statute allows for a claim of disparate impact. Despite the established practice of applying disparate impact in fair lending, in April 2025, a presidential executive order denounced disparate impact liability as a threat to ``meritocracy'' and ``equality of opportunity'' \citep{ExecOrder2025Meritocracy}. Despite these attacks, it seems likely that any changes to the applicability of the disparate impact doctrine requires legislative intervention (in the case of FHA) or at least changes in the implementing regulation (in the case of ECOA).} Disparate treatment deals with the direct conditioning of a credit decision on a protected characteristic. Disparate impact covers cases in which a facially neutral rule has an impermissible disparate effect on protected groups. 

A disparate impact case typically follows the three-part burden-shifting framework originally developed in the Title VII employment discrimination context. First, the plaintiff must make an initial showing that a practice resulted in a disparate outcome for a protected group \citep{gillis2021input,usc_2000e2k,albemarle_paper_v_moody_1975}.  Once a plaintiff has established the disparate outcome and its cause, the burden shifts to the defendant to demonstrate that there was a ``business justification,'' sometimes also referred to as ``business necessity,'' for the policy that led to the disparity \citep{gillis2025price}. If the defendant satisfies this burden, the plaintiff must show that the business interest could be achieved with a less discriminatory alternative ~\citep{black2023less,gillis2024operationalizing}.

\subsubsection{Federal UDAP} The other legal framework we discuss, the prohibition on unfair, deceptive, and abusive acts or practices, originated in Section 5 of the Federal Trade Commission Act \citep{FTCA1914}, which covers individuals and entities involved in interstate commerce. Section 1031 of the Dodd-Frank Act \citep{CFPA2010}, enacted in 2010, restricts providers of consumer financial products or services from engaging in unfair, deceptive, and/or abusive acts or practices.  One way in which this prohibition differs from the FTC Act prohibition is the additional prohibition of ``abusive'' acts or practices. For simplicity, we use the term ``UDAP'' to refer collectively to the FTC Act and Dodd-Frank provisions, even though Dodd-Frank's provision includes an ``abusive'' prong and is often referred to as ``UDAAP''.  Another important difference is that the Federal Trade Commission's (FTC) jurisdiction also covers a broader range of consumer protection issues than financial products and services alone, including advertising and marketing practices across many industries.  The principles of ``unfair'' and ``deceptive'' practices in the Dodd-Frank Act are similar to those under Section 5 of the Federal Trade Commission Act \citep{CFPB2022}.

An \textbf{unfair} act or practice has three elements. First, it must cause or be likely to cause ``substantial injury'' to consumers ~\citep{FTCUnfairness1980,FTCA1994}. This typically means financial harm but can also include other types of harm such as unwarranted health and safety risks. Second, the harm must ``not be reasonably avoidable'': If consumers have a free and informed choice and they choose a course of action that results in harm, then the harm is not ``unfair.'' Finally, the injury must ``not [be] outweighed by countervailing benefits'' to consumers or competition. This ensures that an act or practice is not deemed ``unfair'' if it produces greater benefit than harm.

An act or practice is \textbf{deceptive} if (1) it misleads or is likely to mislead a consumer; (2) the consumer’s interpretation is reasonable under the circumstances; and (3) the representation is material \citep{FTCDeception1983}. 

Finally, \textbf{abusive} conduct is defined under Dodd-Frank as an act or practice that materially interferes with the ability of a consumer to understand the terms of a product or service or that takes unreasonable advantage of the consumer’s lack of understanding, the inability of the consumer to protect their interests or the reasonable reliance by the consumer on the product or service provider to act in the interests of the consumer \citep{CFPA2010}.

\subsubsection{State UDAP} All fifty states, the District of Columbia, Puerto Rico, Guam, and the Virgin Islands have statutes akin to the Federal Trade Commission Act\footnote{Few explicitly protect against abusive practices, so we also use the shorter acronym of UDAP to refer to state laws \citep{nclc_unfair_deceptive_acts_practices_2021}.} \citep{nclc_unfair_deceptive_acts_practices_2021}. State UDAP laws generally expand consumer protection powers to state governments \citep{pridgen2016dynamic} and, with the exception of Puerto Rico, provide consumers with a private cause of action \citep{nclc_unfair_deceptive_acts_practices_2021}. Although the text of these statutes varies considerably \citep{pridgen_consumer_law_2020}, they typically serve similar functions.

The typical state UDAP empowers the state Attorney General to seek injunctions against unlawful conduct and to seek civil penalties, including restitution \citep{nclc_maps_how_well_states_protect_consumers_2018}. To lower barriers for private litigants, many states permit recovery of attorney’s fees and authorize punitive, multiple, or minimum damages, while applying lower burdens of proof \citep{nclc_maps_how_well_states_protect_consumers_2018,nclc_unfair_deceptive_acts_practices_2021,pridgen2016dynamic,pridgen_consumer_law_2020}.

Over time, however, industry lobbying has eroded the strength of many state UDAP laws \citep{gilles2023private}. Several sectors enjoy categorical exemptions: seven states exclude most lenders and creditors, fourteen exempt utilities, and twenty-one exempt insurance companies \citep{nclc_maps_how_well_states_protect_consumers_2018}. Other states have narrowed the effectiveness of private rights of action: nine limit protections against lending and insurance companies, five prohibit recovery of attorney’s fees, and two even require unsuccessful plaintiffs to pay defendants' attorney's fees \citep{nclc_maps_how_well_states_protect_consumers_2018}. Mississippi has adopted so many restrictions that it has “effectively rescinded the private right of action” \citep{gilles2023private}.

Despite these imitations, state UDAPs remain important consumer protection tools. They continue to apply to billions of transactions annually \citep{nclc_maps_how_well_states_protect_consumers_2018} and, because of their broad mandate, allow both Attorneys General and consumers to respond quickly to new forms of fraud \citep{nclc_unfair_deceptive_acts_practices_2021}.\footnote{For example, in as early as 2020 consumers used California’s Unfair Competition Law to bring a cryptocurrency exchange platform to court \citep{crypto_asset_v_hoard_2020}, and in 2022 California State regulators used UDAP authroity issue a desist and refrain order to another cryptocurrency exchange platform \citep{dfpi_gmo_global_desist_refrain_2022}.} Even as narrowed, state UDAPs largely function as an important remedy against emerging forms of consumer harm \citep{nclc_unfair_deceptive_acts_practices_2021}.

\subsection{Targeting Discrimination using UDAAP} 

Both the FTC and the CFPB have recently suggested that UDAP's prohibition, though traditionally focused on consumer-protection misconduct rather than discrimination \textit{per se}, can also be deployed against discriminatory conduct. The FTC's case against \textit{Passport} \citep{FTCPassport2022} was the first enforcement action to challenge discriminatory conduct under the UDAP ``unfairness'' authority. In March 2022 the CFPB announced its intention to bring UDAAP enforcement actions in the context of discriminatory lending, ``including in situations where fair lending laws may not apply'' \citep{cfpb_targets_unfair_discrimination_2022}.\footnote{The announcement highlighted the updated UDAAP manual's statement that UDAAP authority for rulemaking, supervision, and enforcement runs in parallel to other laws, such as ECOA. The announcement came as part of a general update to the UDAAP supervisory guidance to examiners, which included guidance on ``the interplay between unfair, deceptive, or abusive acts or practices and other consumer protection and antidiscrimination statutes'' \citep{CFPB2022}.} However, the CFPB's attempt to embed discrimination within its UDAAP supervisory manual was vacated by a federal court in 2023 \citep{ChamberCFPB2023}, and the Bureau abandoned its appeal in 2025 \citep{chamber_v_cfpb_2025}, leaving the scope of UDAAP in this area uncertain.

The use of the UDAP prohibition for antidiscrimination enforcement has been controversial \citep{ICBA2022}. In his dissent in the \textit{Passport} case, Commissioner Philips argued that the FTC Act is not an antidiscrimination statute, citing the absence of the disparate impact doctrine in its text \citep{phillips_dissent_ftc_passport_oct2022}.
Nonetheless, the majority in \textit{Passport} embraced the unfairness-based discrimination theory, a position foreshadowed by previous FTC statements and supported in a growing body of scholarship \citep{HayesSchellenberg2021,Herrine2023,SelbstBarocas2023,Sovern2024}.


When UDAP authority overlaps with existing discrimination laws, such as ECOA, an important question is whether conduct that would not be deemed discriminatory under ECOA can be considered discriminatory under the UDAP prohibition and vice-versa. This question is particularly contentious when considering whether UDAP, and specifically ``unfairness,'' also covers cases of disparate impact.  The FTC \textit{Passport} case suggests an approach to UDAP discrimination enforcement that closely tracks disparate impact enforcement under ECOA. 

Understanding how DI and UDAP relate to one another is particularly important in the context of fair lending, to which both the FTC UDAP rule and the CFPB UDAAP rule apply, given that Dodd-Frank also covers lenders. Moreover, first attempts to apply UDAAP to discriminatory conduct have happened primarily in the context of fair lending. We argue that, contrary to many regulatory and academic interpretations \citep{FTCPassport2022}, the unfairness prong of UDAP should not be treated as a doctrinal mirror of DI. Instead, it offers a distinct normative and analytical framework, and one that warrants independent attention in the evaluation of algorithmic lending practices.

\section{Operationalizing Legal Standards: A Technical Implementation of DI and UDAAP}
\label{sec:tehcnical}

In this section, we offer a technical implementation of legal standards that have not yet been formalized in algorithmic contexts. We consider the case of a firm training a predictive model for business purposes, where higher accuracy increases the firm’s payoff but may also generate disparities for protected groups. For example, a lender might train a model to predict default risk, where the chosen model produces disparities for men and women. Taking such a baseline model as given, we ask whether the lender would face liability under discrimination law for deploying it, and whether the law would require adoption of an alternative model that reduces those disparities.

We examine three settings for our technical implementation, each involving a distinct prediction task and protected group. Using simulations of a hypothetical firm, we translate the DI and UDAP doctrines into operational criteria that can be applied to evaluate model behavior and compliance. While prior work has explored how DI may be formalized in technical frameworks \citep{gillis2021input,gillis2024operationalizing,caro2023modernizing,black2023less}, to our knowledge, this is the first paper to develop a comparable technical implementation of UDAP. 

Importantly, this technical demonstration relies on a series of assumptions and informed judgments about how key elements of both UDAP, specifically the ``unfairness'' prong, and DI doctrine are to be interpreted. Our goal is \emph{not} to claim that these are necessarily the correct technical interpretations, but rather to highlight the dimensions that require regulatory guidance and clarification. Through our technical implementation of the two doctrines, we identify key points of convergence and divergence that have implications for how algorithmic systems are assessed under each legal standard, and how they manifest under one set of reasonable interpretations. These differences reflect the conditions under which disparities are scrutinized and how they are justified. While the doctrines differ in their wording, our analysis takes a functional perspective, comparing how each element operates in practice and how those elements align or diverge across the two frameworks.


\subsection{Implementation Details}

We examine how the ``unfairness'' doctrine under UDAP and the DI doctrine respond to disparities arising from algorithmic predictions in consumer finance settings. To do so, we consider three stylized prediction tasks that reflect different stages of the credit decision pipeline: (1) predicting income, which may inform assessments of repayment capacity; (2) predicting creditworthiness, as a proxy for risk evaluation; and (3) approximating lending decisions by predicting loan approval outcomes. While each prediction task serves a distinct function, all may yield disparities across protected groups, such as differing approval rates for white and Black applicants, even when they improve overall predictive accuracy.

Our analysis centers on two questions: First, do disparities arising from these models trigger scrutiny under UDAP or DI doctrines, thereby obligating the lender to consider alternative models? Second, what kinds of alternatives would be deemed acceptable under each doctrinal framework? By holding the empirical setting fixed and varying the doctrinal lens, we clarify how legal standards shape the assessment of fairness and model choice.


\subsubsection{Data.}
We perform our experiments on the following datasets: Adult Income~\citep{kohavi1996scaling},\footnote{UCI Adult Income dataset, derived from the 1994 Census; accessed via OpenML at \href{https://www.openml.org/search?type=data\&status=active\&id=43898}{OpenML \#43898}.} UCI German Credit~\citep{gromping2019german},\footnote{{Originally released as German Credit data via Hofmann/Statlog in the 1990s; later corrected and documented by \citet{gromping2019german}; dataset available via OpenML at \href{https://www.openml.org/search?type=data\&status=active\&id=44096\&sort=runs}{OpenML \#44096}.}} and HMDA~\citep{gillis2021input,black2024d}.\footnote{Public HMDA data (Suffolk County, MA) from the FFIEC data browser: \href{https://ffiec.cfpb.gov/data-browser/data/2024?category=counties\&items=25025}{FFIEC 2024}} The Adult Income dataset is a subset of the publicly-available 1994 US Census with the target variable being a binary outcome that indicates whether an individual's annual income exceeds \$50,000.
The German Credit Dataset contains individual financial attributes and uses a binary target indicating whether the applicant is classified as a ``good'' credit risk. The Home Mortgage Disclosure Act (HMDA) contains mortgage application data and indicates whether a mortgage loan was approved in the binary outcome. We use a subset of HMDA data from Boston (Suffolk County, MA) \citep{munnell1996mortgage}. The protected attribute for the UCI German Credit dataset is sex, while for the Adult and HMDA datasets it is race (operationalized as white versus non-white). The Adult and UCI German credit datasets were accessed using OpenML and the HMDA dataset was retrieved directly from the Federal Financial Institutions Examination Council (FFIEC) portal. All datasets are publicly available.
While many of these datasets do not correspond exactly to a task relevant to consumer credit (e.g. predicting loan outcome), they are common publicly available datasets in the fairness and machine learning literature~~\citep{black2024d}, with inputs that closely reflect information considered in consumer credit tasks.

\subsubsection{Generating Models.}

For each dataset, we train multiple configurations of five commonly used classifiers: Logistic Regression, Random Forest, Decision Tree, SVM, and XGBoost. We also train variations of these models using Fairlearn~~\citep{bird2020fairlearn}, a package used to implement disparity mitigations. From the Fairlearn package, we use the following disparity reduction techniques: adversarial debiasing~~\citep{zhang2018mitigating}, fair reductions~~\citep{agarwal2018reductions}, and postprocessing~~\citep{hardt2016equality} methods. 
For each type of classifier, we vary the feature set and certain hyper-parameters to generate a wide range of models. Each choice of model and feature set will lead to a different search over a set of hyper-parameters. To model potential decision-making scenarios, we keep only the models within the top 5\% of accuracy for each search. Once we generate the models from each of these wide searches for each dataset, we only keep the models that result in a maximum 10\% accuracy drop from the most accurate model found throughout the entire search. This model search results in between 250 and 464 models, depending on the dataset. Further training details are in Appendix~\ref{sec:appendix}. 

\subsubsection{Baseline models.} We choose our baseline models post hoc for illustrative purposes. For the Adult and German Credit datasets, we select the highest-accuracy models identified during the search process. For the HMDA demonstration, we select a model with accuracy comparable to the best-performing model but not on the accuracy–fairness frontier. This choice reflects the reality of model search. Maximizing for accuracy alone rarely yields a model that also performs well on fairness metrics, due to underspecification~~\citep{d2020underspecification}. In practice, one is unlikely to incidentally discover strong performance on an axis that was not explicitly optimized~~\citep{d2020underspecification}. Yet models with similar accuracy often exist that achieve meaningful gains in fairness (or in other desiderata) with little sacrifice in performance~~\citep{black2022model, rodolfa2021empirical}. Moreover, explicitly searching for models with low disparity can drive a more thorough search process that uncovers both more accurate and more fair models. As we show in \autoref{sec:appendix}, some of our most accurate models emerged from using Fairlearn’s disparate impact mitigation techniques.

\subsubsection{Metrics of interest.}

\begin{figure}
    \centering
    \includegraphics[width=0.75\textwidth]{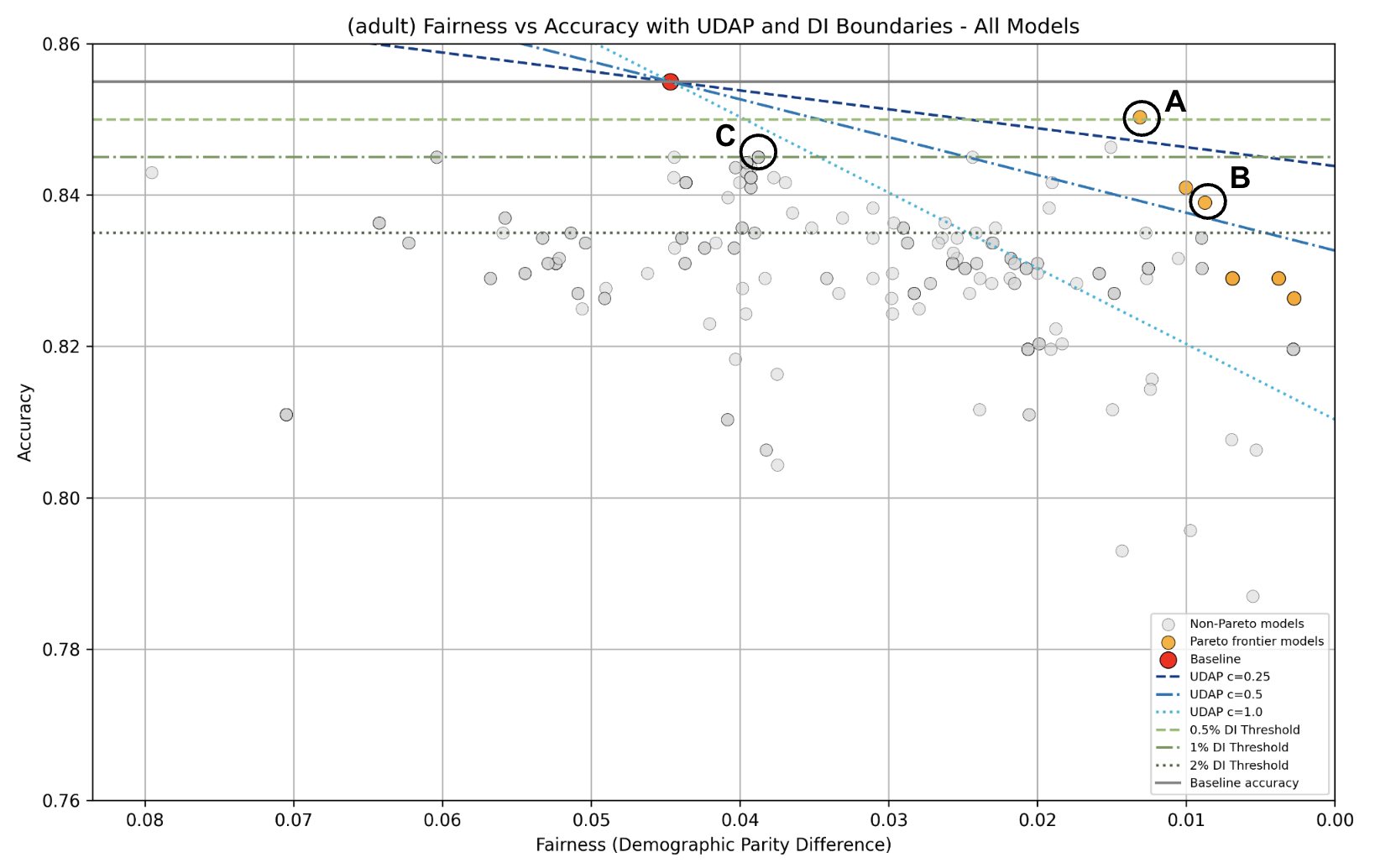}
    \caption{UDAP vs DI Analysis on Adult Data. On the X axis is demographic disparity \emph{decreasing} in severity from left to right, and on the y axis is increasing accuracy. Models on the Pareto frontier are in yellow and the baseline model is denoted in red, but it is also on the Pareto frontier. 
    We note varying possible cut-offs for considering alternatives under DI and UDAAP doctrines, in blue and green. Models that we discuss in detail for legal comparison are noted with letters "A", "B", and "C".}
    \label{fig:adult}
\end{figure}

\begin{figure}
    \centering
    \includegraphics[width=0.75\textwidth]{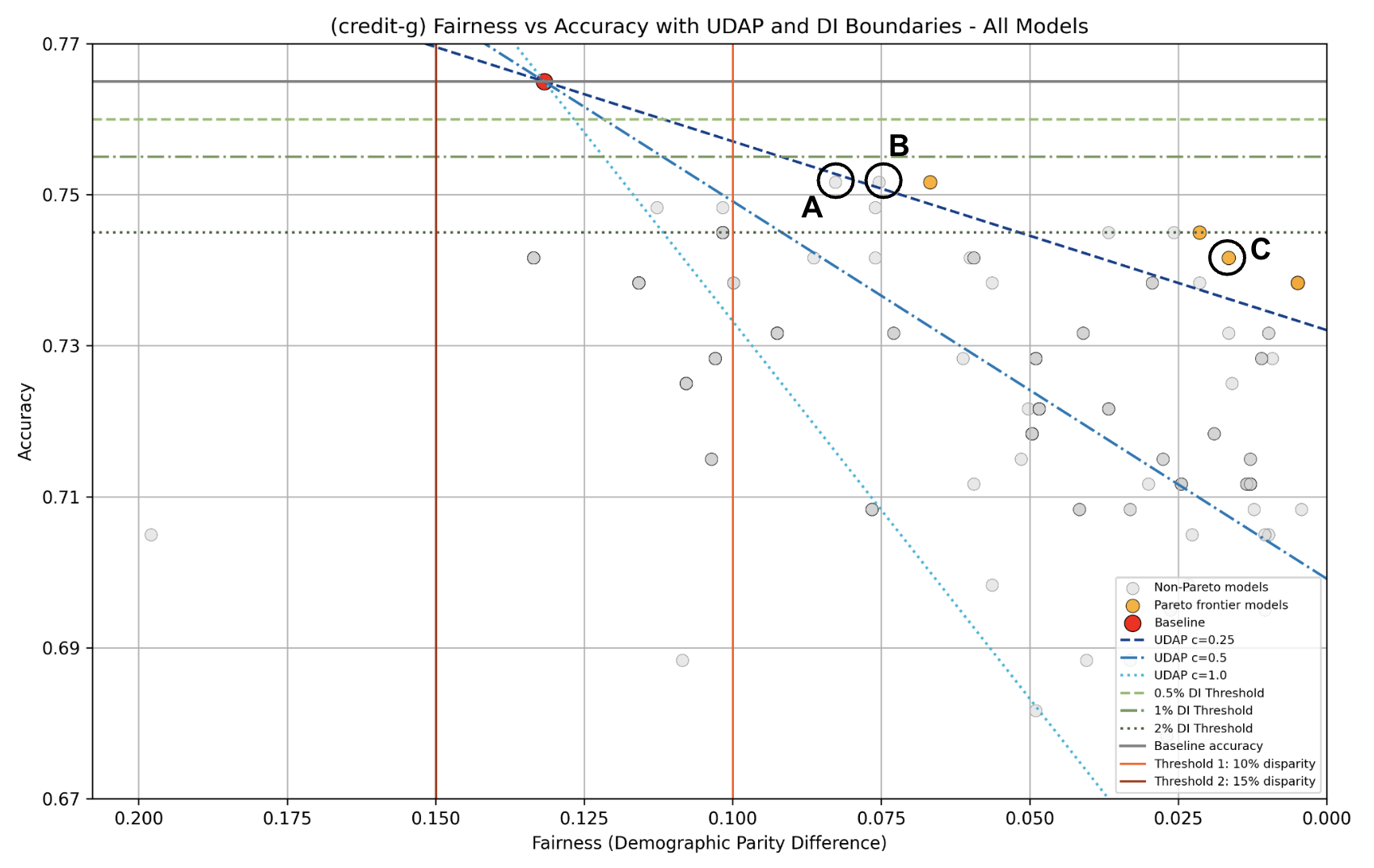}
    \caption{UDAP vs DI Analysis on German Credit Data. 
    On the X axis is demographic disparity \emph{decreasing} in severity from left to right, and on the y axis is increasing accuracy. Models on the Pareto frontier are in yellow and the baseline model is denoted in red, but it is also on the Pareto frontier. We note varying possible cut-offs for considering alternatives under DI and UDAP doctrines, in blue and green. Models that we discuss in detail for legal comparison are noted with letters "A", "B", and "C". We also note different potential thresholds for triggering a DI or UDAP claim through the vertical lines in orange and red.}
    \label{fig:g-credit}
\end{figure}

\begin{figure}
    \centering
    \includegraphics[width=0.75\textwidth]{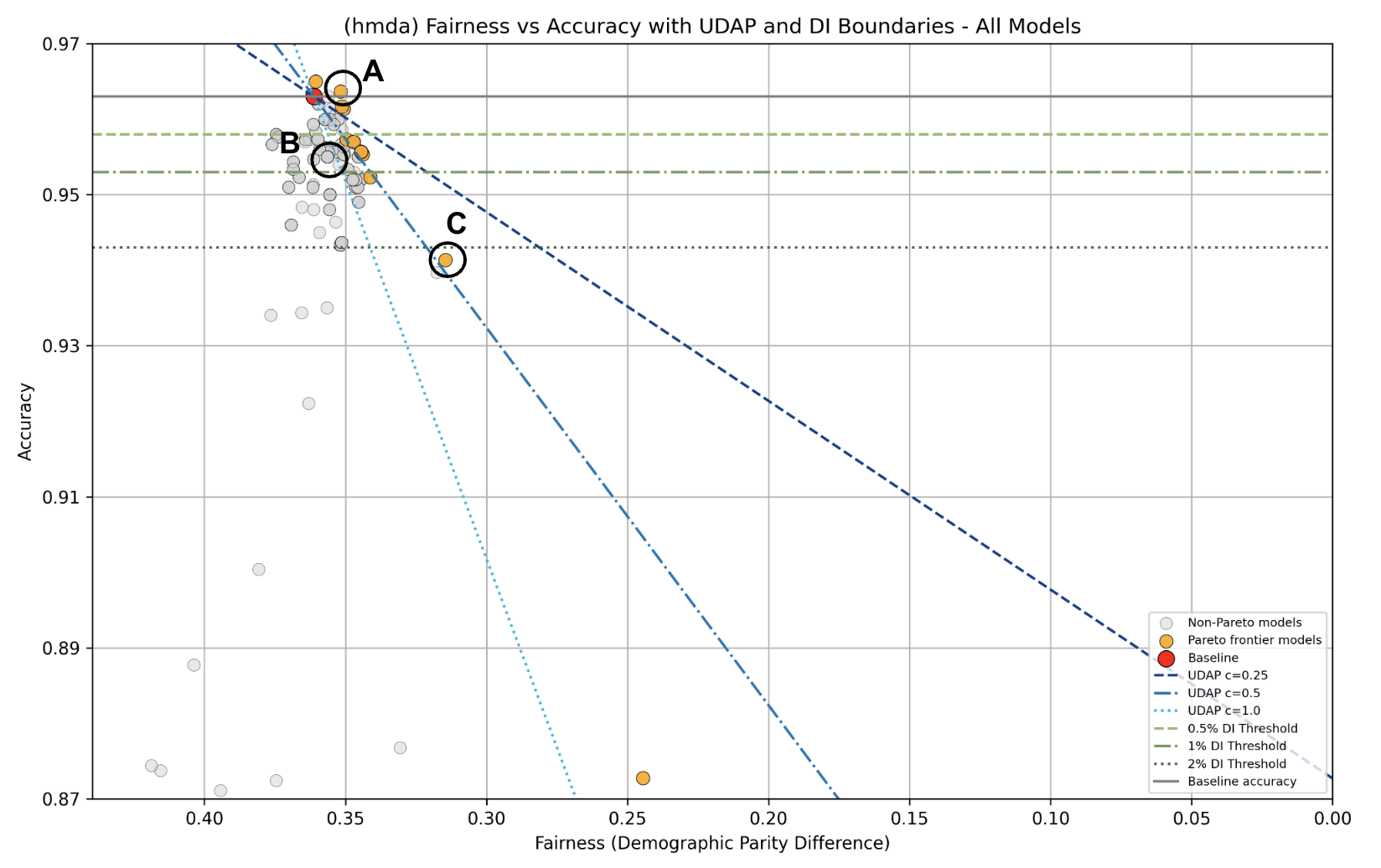}
    \caption{UDAP vs DI Analysis for finding alternatives on HMDA. On the X axis is demographic disparity \emph{decreasing} in severity from left to right, and on the y axis is increasing accuracy.  Models on the Pareto frontier are in yellow and the baseline model is denoted in red, and it is \emph{not} on the Pareto frontier. We note varying possible cut-offs for considering alternatives under DI and UDAP doctrines, in blue and green. Models that we discuss in detail for legal comparison are noted with letters "A", "B", and "C". }
    \label{fig:g-HMDA}
\end{figure}

Once we have our models, we measure them along two dimensions for comparison: accuracy and fairness. To do so, we use one common hold-out per dataset across all models we train. For \textbf{accuracy}, we measure the percentage of correct predictions. For \textbf{fairness}, we calculate the demographic disparity~~\citep{verma2018fairness} of the protected group, i.e. the difference in positive prediction (approval) rates. For example, in the HMDA dataset we look at the difference in approval rates for white and non-white borrowers as our fairness metric. We visualize results by plotting each model as a point in accuracy-fairness space (see \autoref{fig:adult}, \autoref{fig:g-credit} and \autoref{fig:g-HMDA}). We denote models on the \textit{Pareto frontier} of accuracy and fairness---i.e. the set of models for which no other model is strictly better on both accuracy and fairness---in yellow. 


\subsection{Mapping Legal Doctrines to Technical Criteria}

This section summarizes the differences between DI and UDAP, with particular emphasis on the ``unfairness'' prong of UDAP. While both doctrines contain unresolved ambiguities, UDAP is especially underdeveloped in its application to discrimination, and our analysis aims to provide a first step toward clarifying its contours in algorithmic contexts. Accordingly, we present one set of interpretations of how the doctrines diverge, while noting where alternative readings may also be plausible.

\autoref{tab:DI-UDAAP} provides an overview of the two doctrines and highlights three core elements that frame our comparison. First, each doctrine has a threshold showing of harm: under DI this is the plaintiff’s \textit{prima facie} burden to establish disparities, while under UDAP it is the requirement to demonstrate ``substantial injury.'' Second, each doctrine sets out conditions under which disparities may be justified: DI allows defendants to invoke a business justification, whereas UDAP treats harms as non-actionable if they are reasonably avoidable by consumers. Third, both doctrines include a mechanism for weighing harms against benefits, which we interpret as the requirement to consider alternatives. In DI, this takes the form of the less discriminatory alternative requirement, while in UDAP an act or practice is not unfair if its benefits outweigh its costs. We consider each of these three elements in turn. 

\begin{table}[htbp]
\centering
\renewcommand{\arraystretch}{1.3} 
\resizebox{\textwidth}{!}{%
\begin{tabular}{|p{4cm}||p{6cm}|p{6cm}|}
\hline\hline
\textbf{} & \textbf{Disparate Impact} & \textbf{UDAAP (unfairness)} \\
\hline\hline
\textbf{Threshold showing of harm} & 
\textit{Prima facie} showing of disparities; ``Practical significance'' &
``Substantial injury'' \\
\hline
\textbf{Justifications} & 
Business justification / necessity &
``Not reasonably avoidable'' \\
\hline
\textbf{Alternatives} & 
Less discriminatory alternative &
Not ``outweighed by countervailing benefits to consumers or to competition'' \\
\hline\hline
\end{tabular}
}
\caption{Comparison of Disparate Impact and UDAAP (unfairness) doctrines.}
\label{tab:DI-UDAAP}
\vspace{-1em}
\end{table}

\subsubsection{Initial showing of harm/disparities.}
One dimension on which the two doctrines can be compared is whether they have the same threshold of harm or disparities for triggering the doctrine. 

Although the disparate impact framework does not impose a formal numerical threshold for the magnitude of disparity required at the \textit{prima facie} stage, regulatory guidance and enforcement practice suggest that a disparity must be both statistically and practically significant. Practical significance assesses the magnitude of the disparity and whether it is meaningful in real-world terms and originates primarily from the requirement in employment disparate impact settings \citep{EEOCUniformGuidelines1978,Tobia2022Disparate}. While courts have been somewhat divided on whether plaintiffs must show that a disparity is practically significant to make out a \textit{prima facie} case of disparate impact, many institutions routinely apply practical significance thresholds in evaluating automated models to decide whether observed disparities merit further analysis. Some commentators have suggested that for a loan approval/denial disparity to be practically significantly adverse, it must have an adverse impact ratio less than $0.90$  \citep{relman2021upstart}.\footnote{Early cases of disparate impact, particularly in the employment setting, emphasized the need for disparities to be ``sufficiently substantial'' \citep{Watson1988}, however, both the 2013 Disparate Impact rule of the Department of Housing and Urban Development (HUD) \citep{HUD2013DisparateImpact}, implementing the FHA, and the Joint Policy Statement on Lending Discrimination \citep{JointPolicy1994}, do not lay down a threshold requirement for the first stage of disparate impact.}

The statutory and regulatory definition of unfairness requires a showing that a practice ``causes or is likely to cause substantial injury to consumers which is not reasonably avoidable by consumers and not outweighed by countervailing benefits to consumers or to competition'' \citep{usc_12_5531c, usc_15_45n}. This is understood to impose three requirements, the first of which is that an act or practice must cause or be likely to cause ``substantial injury'' to consumers. The FTC's longstanding unfairness policy statement clarifies that the injury must be significant and not ``trivial or merely speculative'' \citep{FTCUnfairness1980}, and courts have generally interpreted substantial injury to include monetary harm or significant risks of concrete harm \citep{neovi2010, orkin1988}. 

UDAP's ``substantial injury'' can also be viewed as acting as a threshold requirement for triggering the doctrine. In comparing unfairness' ``substantial injury'' requirement to DI's ``practical significance'' requirement (assuming that such a requirement exists for fair lending disparate impact), it is unclear which of the two requirements is stricter in terms of the magnitude of disparities that must be demonstrated. It may be the case that UDAP's notion of substantial injury requires a higher threshold of disparity than DI's ``practically significant,'' however, more guidance on the precise threshold of these requirements is needed for an accurate comparison. 

\paragraph{Empirical Demonstration} In ~\autoref{fig:g-credit}, we demonstrate two potential thresholds for what might constitute a meaningful disparity on the German Credit dataset---one at a 10\% disparity in the selection rate between male and female borrowers and one at a 15\%. Consider that the 10\% threshold meets the DI requirement of statistical and practical significance, meaning that the model in red is sufficient to trigger the DI doctrine by meeting the \textit{prima facie} showing of disparities under the first stage of DI.  If the ``substantial injury'' is a more stringent requirement than under DI, we may say that unfairness under UDAP is not triggered until the 15\% disparity threshold. Then the baseline model highlighted in red would not be considered liable under UDAP. One interpretation might be that, at very small levels of discrimination, the relationship between the doctrines could reverse: a disparity too limited to meet DI's requirement of practical significance might still impose substantial injury on a few consumers that is not offset by countervailing benefits, thereby triggering UDAP's unfairness standard.


\subsubsection{Justifications.}
Both UDAP and DI provide some avenue to justify disparities created by a firms's policies. As noted above, under DI's burden-shifting framework originally developed in the Title VII employment discrimination context, after the plaintiff shows that a practice resulted in a disparate outcome for a protected group, the burden shifts to the defendant to demonstrate that there was a ``business justification,'' sometimes also referred to as ``business necessity,'' for the policy that led to the disparity \citep{gillis2025price}. 

UDAP does not have an equivalent of DI's ``business justification,'' however UDAP's unfairness second element, defining unfair practices as those that are ``not reasonably avoidable'' by consumers,  could also be thought of as providing an avenue of defense and justification for the disparities created.\footnote{While the third element of unfairness which looks to consider countervailing benefits, may also form a type of justification, we consider it closer to  DI's less discriminatory alternative prong, as these elements provide more of a balancing function while justifications act more as binary permissions. Meaning, if these justifications are met, they justify disparities without an explicit form of cost-benefit analysis.} While this could be used by firms to argue that harm could have been avoided through available channels, both scholars and regulators have suggested that by definition, discrimination is not reasonably avoidable, since an individual cannot change immutable characteristics such as race to achieve a better outcome. 

Until recently, the CFPB's UDAAP Manual explicitly stated that ``consumers cannot reasonably avoid discrimination.''\footnote{Until Sept. 2023 \citep{CFPB2022}. Also see discussion in \autoref{sec:legal} on the current status of the Examination Manual}  That language has since been removed, but one can still argue that discriminatory AI practices are effectively unavoidable. Algorithmic systems are opaque both to companies and applicants, and their growing homogeneity and ubiquity make it difficult for consumers to identify or evade discriminatory effects.\footnote{The vacated CFPB UDAAP Manual emphasized that the relevant question is not whether a consumer could have made a better choice, but whether an act or practice hinders decision-making. It further noted that consumers are not expected to take impractical measures, such as hiring independent experts or filing lawsuits in every instance of harm, to avoid injury \citep{CFPB2022}.} Because of this opacity~~\citep{gillis2025price,rudin2019stop}, expecting consumers to recognize discrimination in algorithmic decision-making and to seek out alternatives may stretch beyond what the justification reasonably allows.

Given that the analysis of justification requirements include specific details of the motivation for a policy and implementation details beyond what we include here,  we do not include this analysis in our technical implementation.


\subsubsection{Alternatives.}
Once a threshold for triggering a discrimination claim is established, and there is a business necessity for the policy or it is not reasonably avoidable, the next question is how each doctrine treats the role of alternatives? Both unfairness and DI provide frameworks for weighing potential benefits to firms and consumers against harmful disparities, but they do so through different lenses. Under unfairness the inquiry is framed as a cost–benefit analysis, asking whether harms are outweighed by benefits. Under DI the focus is on whether an alternative less discriminatory provides  sufficient ability to meet business needs. Depending on how these standards are interpreted, each doctrine can validate and reject models that the other would not. In what follows, we outline key ways the two doctrines may diverge in their treatment of alternatives, while emphasizing that both remain under-specified and would benefit from regulatory clarification.

\textit{DI Alternatives:} 
The third stage of disparate impact analysis provides that even if a challenged policy is supported by a legitimate business necessity, it may not be used if a less discriminatory alternative can achieve the same business objective. The central question, however, is how closely an alternative must replicate the baseline policy’s performance in order to qualify as a true alternative.

Regulatory and judicial formulations are notably vague. There are several different descriptions of this requirement, from ``equally effective''~~\citep{interagency_fair_lending_2009_Appendix} to ``approximately equally effective''~~\citep{interagency_fair_lending_2009_Appendix} to ``serve the same purpose with less discriminatory effect,''~~\citep{interagency_fair_lending_2009} and even more different interpretations of what each of these means in context.\footnote{Importantly, while the Civil Rights Act of 1991 codified DI under Title VII (employment discrimination) and overturned the restrictive standard of ``equally effective'' in the Supreme Court decision \textit{Wards Cove Packing Co. v. Atonio} \citep{WardsCove1989}, DI in credit arises instead from agency interpretation of ECOA, most notably in Regulation B \citep{RegB} and its Official Interpretations. These interpretations recognize the ``effects test'' but do not specify what counts as a less discriminatory alternative. Courts and regulators have varied in their formulations, sometimes referring to alternatives that are ``equally effective,'' ``approximately equally effective,'' or ``serve the same purpose with less discriminatory effect'' to achieve the business objective~\citep{black2023less,interagency_fair_lending_2009_Appendix,interagency_fair_lending_2009}. HUD’s FHA disparate impact rule provides a parallel illustration: in adopting the 2013 Rule (and reaffirming it in 2023 after rescinding the 2020 version), HUD rejected the view that an LDA must be ``equally effective'' or ``at least as effective'' as the challenged practice, instead requiring only that the alternative be ``supported by evidence'' and ``not be hypothetical or speculative'' \citep{HUD2013DisparateImpact}.}

Emerging evidence suggests that financial institutions adopt a variety of practical approaches in considering less discriminatory alternatives~\citep{black2025working}. Few institutions apply rigid rules. Instead, decisions are often the product of internal deliberation among multiple stakeholders. In practice, several recurring thresholds appear: (a) alternatives within one to two percentage points of the baseline model on the chosen business performance metric (such as accuracy), (b) alternatives that fall within the confidence interval of the baseline model's performance~\citep{relman2021upstart}, or (c) alternatives that, even if somewhat less accurate than the current model, still outperform an earlier version of the institution’s system. Across these approaches, practitioners emphasize the need for clearer regulatory guidance on what constitutes a sufficient alternative.


\textit{UDAP Alternatives:} 
The third prong of unfairness provides that a practice is unlawful if the resulting injury is ``not outweighed by countervailing benefits to consumers or to competition.'' Unlike DI, there is no formal element addressing alternatives. Still, this cost-benefit inquiry can be understood as a framework for evaluating whether a less harmful model should replace the baseline policy.

As others have noted \citep{SelbstBarocas2023}, the scope of the cost-benefit analysis is decisive. A broad framing, such as treating the overall efficiency gains from deploying an AI system as the relevant benefit, would allow a wide range of practices to survive scrutiny, even where they impose significant group-level harms. For this reason, we adopt a narrower approach. The relevant benchmark is the specific practice at issue, the population exposed to harm, and only those benefits strictly necessary to achieve the identified business objective.

Consistent with this approach, our analysis deliberately limits the scope of benefits to predictive accuracy, weighed against group-level disparities. Under this narrow lens, an alternative model is preferable if, relative to the baseline, it reduces disparities without an offsetting loss in accuracy, or if any loss in accuracy is outweighed by fairness gains. Broader considerations, such as generalized claims of lower prices or wider product availability \citep{FTCA1914} fall outside the frame we adopt here.

Even with a narrowed analysis, important questions remain about how to trade off accuracy and disparity. One issue is how much weight to assign group disparities relative to performance measures. Another is whether the tradeoff should be linear: for example, requiring two points of disparity reduction for each point of accuracy lost or whether more complex, non-linear relationships should govern when a model qualifies as a viable alternative. These design choices remain unresolved by current regulatory guidance.

\paragraph{Empirical Demonstration.} 

Figures~\ref{fig:adult}, \ref{fig:g-credit}, and \ref{fig:g-HMDA} illustrate how different interpretations of DI and UDAP would accept or reject candidate models. For DI, the green dotted lines represent alternative thresholds of acceptable performance loss (ranging up to 2\% degradation). Any model above a given green line is permissible under that DI threshold. For unfairness under UDAP, the blue lines represent potential linear tradeoffs between accuracy loss and disparity reduction, ranging from the strict requirement that each point of accuracy loss be offset by four points of disparity reduction (darkest line) to a one-for-one tradeoff (lightest line). Models above a given blue line satisfy that interpretation of the unfairness standard.  
As we can see in Figures~\ref{fig:adult}, \ref{fig:g-credit}, and~\ref{fig:g-HMDA}, both unfairness and DI admit a variety of models as alternatives, and they each admit models that the other would not. 
On the \textit{Adult} dataset (\autoref{fig:adult}), model B fails under all but the most lenient DI threshold but qualifies under the $c=1, 0.5$ unfairness standards. By contrast, model C qualifies under lenient DI thresholds but not under any UDAP tradeoff, while model A is acceptable under most interpretations of both doctrines.  

A similar pattern appears in the \textit{German Credit} dataset (\autoref{fig:g-credit}). Model C qualifies under all UDAP thresholds but under none of the DI thresholds. Models A and B highlight a distinctive feature of the UDAP framework: although nearly identical in accuracy, only model B satisfies the $c=0.25$ tradeoff because it achieves a greater fairness gain.  

In the \textit{HMDA} dataset (\autoref{fig:g-HMDA}), the divergence persists. Model B qualifies under DI but not under any UDAP threshold, while model C qualifies under two UDAP bounds but not DI. Model A is notable as it improves both accuracy and fairness simultaneously, making it acceptable.  

These demonstrations underscore that DI and unfairness under UDAP can point to different sets of acceptable alternatives. DI permits any model within a specified accuracy threshold (e.g., model C in \autoref{fig:adult}, which reduces disparity by 0.75\% while remaining within a 1\% accuracy loss). UDAP, by contrast, can justify larger decreases in accuracy if accompanied by substantial fairness gains (e.g., in \autoref{fig:g-credit}, models toward the right nearly eliminate disparity at the cost of a 4.5\% accuracy decrease). In short, DI cabins alternatives by accuracy thresholds, while UDAP flexibly trades accuracy for disparity reduction, depending on how the cost-benefit rule is specified.

\section{Extensions and Emerging Frontiers in Consumer Protection}
\label{sec:future_guidance}


\subsection{Importance of State UDAP Laws}  

At a time when federal enforcement of UDAP and DI is being scaled back, state UDAP laws may represent the most meaningful avenue for consumer protection. Three weeks into the second Trump administration, the CFPB ordered staff to cease investigative work \citep{rugaber_trump_consumer_protection_cease_2025}. Anticipating this shift, the Bureau itself released a report just before the inauguration, urging states to strengthen their UDAP statutes \citep{cfpb_strengthening_state_level_consumer_protections_2025}. Since then, state attorneys general have increasingly stepped into the gap: for example, the New York Attorney General has pursued enforcement actions abandoned by the CFPB \citep{nyag_james_sues_zelle_2025}, and financial commentators have warned that responsibility for consumer protection is now shifting to the states \citep{sundheim_future_of_udaap_2025}.  

The stakes are significant. State UDAP statutes are often broader and more flexible than federal law, and the Dodd-Frank Act authorizes state attorneys general to enforce federal UDAP provisions directly \citep{usc_5552a1}. This means that even as federal agencies retreat, states retain the tools to preserve a baseline of consumer protection. In the coming years, effective enforcement against harmful or discriminatory AI use may depend on whether and how aggressively states use these powers.

\subsection{Deceptive and Abusive Acts and Practices}  

Our analysis in \autoref{sec:tehcnical} focused on the ``unfairness'' prong of UDAP and how it applies to traditional AI systems. In this section, we turn to UDAP's prohibition on ``deceptive'' and ``abusive'' acts or practices,
with particular attention to GenAI and large language models (LLMs).  We begin by examining consumer harms arising from model \textit{hallucinations}~~\citep{huang2025survey} and then consider how discriminatory concerns may also be addressed through the prohibitions on deception and abuse.

\subsubsection{Harms under the deceptive and abusive prongs}

The UDAP prohibition on deceptive conduct under the FTC Act and Dodd-Frank, as well as the prohibition on abusive conduct under Dodd-Frank (UDAAP), may provide ways to challenge harms arising from LLM hallucinations~~\citep{huang2025survey, zhang2025siren}---i.e. when LLMs generate seemingly plausible but factually unsupported content. As discussed in \autoref{sec:legal}, conduct is ``deceptive'' if it is misleading or likely to mislead a reasonable consumer on a material matter. Certain forms of LLM hallucination plausibly meet this definition.  

Foundation models developed by OpenAI, Google, Anthropic, and others are now widely used by consumers for advice across diverse domains, ranging from recipes to financial planning~~\citep{chatterji2025people,creditkarma2025finai}. Increasingly, firms in sectors with direct implications for consumer safety and welfare, such as tax preparation~~\citep{fowler2024turbotax}, healthcare, and insurance~~\citep{microsoft2024transparencynote}, are integrating Generative AI chatbots into their consumer-facing services. This high-stakes deployment is especially concerning in light of recent studies showing that LLM responses to questions with substantial personal consequences, such as medical or financial queries, are frequently incorrect or misleading~~\citep{birkun2023large,agarwal2024medhalu,fowler2024turbotax}. Reliance on such advice can result in serious physical, financial, or other harm.  

Many providers attempt to mitigate these risks through disclaimers embedded in APIs or user interfaces. For example, ChatGPT warns that ``ChatGPT can make mistakes. Check important info''~~\citep{openaiChatGPT2025}, while Microsoft’s Azure healthcare agent states ``This message is generated by AI and does not provide or replace professional medical advice. Make sure it is accurate and appropriate before relying on this response''~~\citep{microsoft2024transparencynote}. Despite their prevalence, it is unclear whether such disclaimers are sufficient to shield firms from liability. Disclaimers may not prevent reasonable reliance, especially when outputs are provided in contexts where consumers expect authoritative advice or when the information appears highly plausible.   

FTC guidance and case law reinforce this concern. The FTC has emphasized that ``pro forma statements or disclaimers may not cure otherwise deceptive messages or practices''~~\citep{FTCA1994}. Enforcement actions and judicial decisions echo this principle. In \textit{FTC v. FleetCor} \citep{FTC_FleetCor2022}, for example, the district court rejected FleetCor’s reliance on fine-print disclaimers, concluding that ``the tiny, inscrutable print of the disclaimers does not cure the net impression'' created by the ads on per-gallon fuel savings. Similar reasoning could apply to GenAI systems: where consumers are induced to rely on AI outputs, particularly in settings with GenAI agents deployed to provide advice, disclaimers are unlikely to negate the overall impression that the system is providing reliable information.  



The abusive prohibition under Dodd-Frank applies only to covered persons and service providers offering consumer financial products or services. Whereas deception targets misleading content, abuse addresses structural imbalances and covers practices that materially interfere with consumer understanding or exploit consumer vulnerabilities. This standard is salient for GenAI hallucinations: a mortgage servicer’s chatbot that misstates repayment terms, or a tax-preparation bot that fabricates deductions, may not only mislead but also exploit consumer reliance—capturing risks that deception alone may not reach.  

\subsubsection{Deceptive and abusive discriminatory concerns}  

The prohibitions on deception and abuse may also complement discrimination law in settings where harms are not strictly allocative. Disparate impact doctrine is typically applied to allocation decisions, such as loan approvals or pricing, whereas GenAI hallucinations may cause consumer injury without directly altering access to a product or service. For example, if an LLM systematically provides more misleading information to users of a protected class, the resulting disparities could be framed as discriminatory (disparate impact) and simultaneously as deceptive or abusive. In such cases, UDAP under the FTC Act, State UDAPs, and UDAAP under Dodd-Frank can offer an alternative approach for addressing discriminatory harms that fall outside traditional allocative contexts.

\section{Conclusion}

This paper has offered the first technical comparison of disparate impact and UDAP in algorithmic settings, translating their doctrinal elements into operational criteria and demonstrating how they might diverge when applied to the same predictive models. By showing where each doctrine converges and where they might depart, particularly in the treatment of thresholds, justifications, and alternatives, we highlight both the promise and the limits of existing legal tools for addressing algorithmic disparities. Our focus has been on discrimination, but we also suggest that UDAP's reach extends beyond allocative harms to encompass other risks from AI, including deception and abuse. In both contexts, the lack of doctrinal clarity underscores the need for regulatory guidance to determine how these standards apply in practice. With federal enforcement uncertain, state UDAP laws and the broader application of deception and abuse prohibitions may become critical tools for safeguarding consumers. Clarifying how these doctrines should apply to AI is therefore not only a technical challenge but also a regulatory necessity, shaping whether law remains responsive to new forms of inequality embedded in technological systems.

\newpage
\bibliographystyle{ACM-Reference-Format}
\bibliography{bib}

@misc{ECOA1974,
  title     = {Equal Credit Opportunity Act},
        author  = {ECOA},
  year      = {1974}
}

@inproceedings{zhang2018mitigating,
  title={Mitigating unwanted biases with adversarial learning},
  author={Zhang, Brian Hu and Lemoine, Blake and Mitchell, Margaret},
  booktitle={Proceedings of the 2018 AAAI/ACM Conference on AI, Ethics, and Society},
  pages={335--340},
  year={2018}
}

@article{birkun2023large,
  title={Large Language Model (LLM)-Powered Chatbots Fail to Generate Guideline-Consistent Content on Resuscitation and May Provide Potentially Harmful Advice},
  author={Birkun, Alexei A and Gautam, Adhish},
  journal={Prehospital and Disaster Medicine},
  volume={38},
  number={6},
  pages={757--763},
  year={2023},
  publisher={Cambridge University Press}
}

@misc{openaiChatGPT2025,
  title        = {ChatGPT},
  howpublished = {\url{https://chatgpt.com/}},
  note         = {Accessed: 2025-09-25},
  year         = {2025},
  organization = {OpenAI}
}

@misc{microsoft2024transparencynote,
  title        = {Transparency Note: healthcare agent service - Copilot Features},
  howpublished = {\url{https://learn.microsoft.com/en-us/azure/health-bot/transparency-note }},
  note         = {Accessed: 2025-09-29},
  year         = {2024},
  month        = oct,
  organization = {Microsoft},
  date         = {2024-10-20}
}

@article{fowler2024turbotax,
  title={TurboTax and H\&R Block now use AI for tax advice. It’s awful},
  author={Fowler, Geoffrey A.},
  journal={The Washington Post},
  year={2024}
}

@article{agarwal2024medhalu,
  title={Medhalu: Hallucinations in responses to healthcare queries by large language models},
  author={Agarwal, Vibhor and Jin, Yiqiao and Chandra, Mohit and De Choudhury, Munmun and Kumar, Srijan and Sastry, Nishanth},
  journal={arXiv preprint arXiv:2409.19492},
  year={2024}
}

@techreport{chatterji2025people,
  title={How People Use ChatGPT},
  author={Chatterji, Aaron and Cunningham, Thomas and Deming, David J. and Hitzig, Zoe and Ong, Christopher and Shan, Carl Yan and Wadman, Kevin},
  year={2025},
  institution={National Bureau of Economic Research}
}

@article{zhang2025siren,
  title={Siren’s Song in the AI Ocean: A Survey on Hallucination in Large Language Models},
  author={Zhang, Yue and Li, Yafu and Cui, Leyang and Cai, Deng and Liu, Lemao and Fu, Tingchen and Huang, Xinting and Zhao, Enbo and Zhang, Yu and Chen, Yulong and others},
  journal={Computational Linguistics},
  pages={1--46},
  year={2025},
  publisher={MIT Press 255 Main Street, 9th Floor, Cambridge, Massachusetts 02142, USA~…}
}

@article{huang2025survey,
  title={A survey on hallucination in large language models: Principles, taxonomy, challenges, and open questions},
  author={Huang, Lei and Yu, Weijiang and Ma, Weitao and Zhong, Weihong and Feng, Zhangyin and Wang, Haotian and Chen, Qianglong and Peng, Weihua and Feng, Xiaocheng and Qin, Bing and others},
  journal={ACM Transactions on Information Systems},
  volume={43},
  number={2},
  pages={1--55},
  year={2025},
  publisher={ACM New York, NY}
}

@article{rudin2019stop,
  title={Stop explaining black box machine learning models for high stakes decisions and use interpretable models instead},
  author={Rudin, Cynthia},
  journal={Nature machine intelligence},
  volume={1},
  number={5},
  pages={206--215},
  year={2019},
  publisher={Nature Publishing Group UK London}
}

@misc{black2025working,
  title={ Less Discriminatory Algorithms On the Ground: An Empirical Account of Fair Lending Programs.},
  author={Emily Black and Logan Koepke and Mingwei Hsu and Miranda Bogen and Solon Barocas},
  year={2025},
  publisher={Working Paper}
}

@article{rodolfa2021empirical,
  title={Empirical Observation of Negligible Fairness---Accuracy Trade-offs in Machine Learning for Public Policy},
  author={Rodolfa, Kit T. and Lamba, Hemank and Ghani, Rayid},
  journal={Nature Machine Intelligence},
  volume={3},
  number={10},
  pages={896--904},
  year={2021},
  publisher={Nature Publishing Group UK London}
}

@inproceedings{black2022model,
  title={Model multiplicity: Opportunities, concerns, and solutions},
  author={Black, Emily and Raghavan, Manish and Barocas, Solon},
  booktitle={Proceedings of the 2022 ACM conference on fairness, accountability, and transparency},
  pages={850--863},
  year={2022}
}

@article{d2020underspecification,
  title={Underspecification Presents Challenges for Credibility in Modern Machine Learning. ArXiv 2011.03395},
  author={D'Amour, A. and Heller, K.A. and Moldovan, D.I. and Adlam, B. and Alipanahi, B. and Beutel, A. and Chen, C. and Deaton, J. and Eisenstein, J. and Hoffman, M.D. and others},
  journal={arXiv preprint arXiv:2011.03395 [cs, stat]},
  year={2020}
}

@misc{creditkarma2025finai,
  title        = {The Rise of Fin-AI: Why Americans Are Trusting Generative AI With Their Wallets},
  howpublished  = {\url{https://www.creditkarma.com/about/commentary/the-rise-of-fin-ai-why-americans-are-trusting-generative-ai-with-their-wallets}},
  note         = {Accessed: 2025-09-29},
  year         = {2025},
  month        = sep,
  organization = {Intuit Credit Karma},
  date         = {2025-09-02}
}

@article{hardt2016equality,
  title={Equality of opportunity in supervised learning},
  author={Hardt, Moritz and Price, Eric and Srebro, Nati},
  journal={Advances in neural information processing systems},
  volume={29},
  year={2016}
}

@inproceedings{agarwal2018reductions,
  title={A reductions approach to fair classification},
  author={Agarwal, Alekh and Beygelzimer, Alina and Dud{\'\i}k, Miroslav and Langford, John and Wallach, Hanna},
  booktitle={International Conference on Machine Learning},
  pages={60--69},
  year={2018},
  organization={PMLR}
}

@techreport{bird2020fairlearn,
    author = {Bird, Sarah and Dud{\'i}k, Miro and Edgar, Richard and Horn, Brandon and Lutz, Roman and Milan, Vanessa and Sameki, Mehrnoosh and Wallach, Hanna and Walker, Kathleen},
    title = {Fairlearn: A oolkit for ssessing and mproving airness in {AI}},
    institution = {Microsoft},
    year = {2020},
    month = {May},
    url = "https://www.microsoft.com/en-us/research/publication/fairlearn-a-toolkit-for-assessing-and-improving-fairness-in-ai/",
    number = {MSR-TR-2020-32},
}

@article{kim2022race,
  title={Race-aware algorithms: Fairness, nondiscrimination and affirmative action},
  author={Kim, Pauline T.},
  journal={California Law Review},
  volume={110},
  pages={1539},
  year={2022},
  publisher={HeinOnline}
}

@misc{FTCA1914,
  title     = {Federal Trade Commission Act},
  author    = {FTCA},
  year      = {1914}
}

@misc{CFPA2010,
  title     = {Consumer Financial Protection Act},
  author    = {CFPA},
  year      = {2010}
}

@misc{CFPB2022,
  author    = {{Consumer Financial Protection Bureau}},
  title     = {Unfair, Deceptive, or Abusive Acts or Practices Manual, Version 3},
  year      = {2022},
  url       = {https://web.archive.org/web/20220316180829/https://files.consumerfinance.gov/f/documents/cfpb_unfair-deceptive-abusive-acts-practices-udaaps_procedures.pdf}
}

@misc{ICBA2022,
  author    = {{Independent Community Bankers of America}},
  title     = {Unfairness and Discrimination: Examining the CFPB’s Conflation of Distinct Statutory Concepts},
  year      = {2022},
  url       = {https://www.icba.org/all-products/product-details/unfairness-and-discrimination},
  note      = {Claiming that additional authority is required to address discriminatory conduct}
}

@misc{HayesSchellenberg2021,
  author    = {Stephen Hayes and Kali Schellenberg},
  title     = {Discrimination is ``Unfair'': Interpreting UDA(A)P to Prohibit Discrimination},
  year      = {2021},
  organization = {Student Borrower Protection Center},
  url       = {https://protectborrowers.org/wp-content/uploads/2021/04/Discrimination_is_Unfair.pdf},
}

@unpublished{Herrine2023,
  author    = {Luke Herrine},
  title     = {Consumer Protection After Consumer Sovereignty},
  year      = {2023},
  note      = {Manuscript, available from the author}
}

@article{SelbstBarocas2023,
  author    = {Andrew D. Selbst and Solon Barocas},
  title     = {Unfair Artificial Intelligence: How FTC Intervention Can Overcome the Limitations of Discrimination Law},
  journal   = {University of Pennsylvania Law Review},
  volume    = {171},
  pages     = {1023--1088},
  year      = {2023}
}

@article{gillis2021input,
  title={The input fallacy},
  author={Gillis, Talia B.},
  journal={Minnesota Law Review},
  year={2021},
}

@inproceedings{gillis2024operationalizing,
  title={Operationalizing the Search for Less Discriminatory Alternatives in Fair Lending},
  author={Gillis, Talia B. and Meursault, Vitaly and Ustun, Berk},
  booktitle={The 2024 ACM Conference on Fairness, Accountability, and Transparency},
  year={2024}
}

@article{caro2023modernizing,
  title        = {Modernizing Fair Lending},
  author       = {Caro, Spencer and Gillis, Talia B. and Nelson, Scott},
  year         = {2023},
  journal      = {SSRN Electronic Journal},
}

@article{black2023less,
      title={The Legal Duty to Search for Less Discriminatory Algorithms}, 
      author={Emily Black and Logan Koepke and Pauline Kim and Solon Barocas and Mingwei Hsu},
      year={2024},
      journal ={arXiv:2406.06817}, 
}

@article{Sovern2024,
  author    = {Jeff Sovern},
  title     = {Is Discrimination Unfair?},
  journal   = {Georgia State University Law Review},
  year      = {2024},
  note      = {Forthcoming},
  url       = {https://papers.ssrn.com/sol3/papers.cfm?abstract_id=4712271}
}

@article{gillis2025price,
  title={"Price Discrimination" Discrimination},
  author={Gillis, Talia B.},
  journal={Harvard Business Law Review},
  volume={15},
  pages={99},
  year={2025},
  publisher={HeinOnline}
}

@misc{WhiteHouse2025EO14281,
  author       = {{White House}},
  title        = {Restoring Equality of Opportunity and Meritocracy},
  year         = {2025},
  month        = apr,
  day          = {23},
  note         = {Executive Order No.~14281},
  howpublished = {\url{https://www.whitehouse.gov/presidential-actions/2025/04/restoring-equality-of-opportunity-and-meritocracy/}}
}

@inproceedings{kohavi1996scaling,
  author    = {Ron Kohavi},
  title     = {Scaling Up the Accuracy of Naive-Bayes Classifiers: A Decision-Tree Hybrid},
  booktitle = {Proceedings of the Second International Conference on Knowledge Discovery and Data Mining (KDD-96)},
  year      = {1996},
  pages     = {202--207},
  url       = {https://www.aaai.org/Papers/KDD/1996/KDD96-033.pdf}
}

@techreport{gromping2019german,
  author      = {Ulrike Grömping},
  title       = {South German Credit Data: Correcting a Widely Used Data Set},
  institution = {Beuth University of Applied Sciences Berlin},
  year        = {2019},
  number      = {Report 2019-004},
  url         = {https://www1.beuth-hochschule.de/FB_II/reports/Report-2019-004.pdf}
}

@inproceedings{black2024d,
  title={D-hacking},
  author={Black, Emily and Gillis, Talia and Hall, Zara Yasmine},
  booktitle={Proceedings of the 2024 ACM Conference on Fairness, Accountability, and Transparency},
  pages={602--615},
  year={2024}
}

@techreport{relman2021upstart,
  title        = {Fair Lending Monitorship of Upstart Network’s Lending Model: Second Report of the Independent Monitor},
  author       = {{Relman Colfax}},
  institution  = {Relman Colfax PLLC},
  year         = {2021},
  month        = {November},
  url          = {https://www.relmanlaw.com/media/publication/1180_PUBLIC%20Upstart%20Monitorship_2nd%20Report_FINAL-2.pdf},
}

@article{pridgen2016dynamic,
  title={The Dynamic Duo of Consumer Protection: State and Private Enforcement of Unfair and Deceptive Trade Practices Laws},
  author={Pridgen, Dee},
  journal={Antitrust Law Journal},
  volume={81},
  pages={911},
  year={2016},
  publisher={HeinOnline}
}

@book{nclc_unfair_deceptive_acts_practices_2021,
  title        = {Unfair and Deceptive Acts and Practices},
  edition      = {11th},
  author       = {{National Consumer Law Center}},
  year         = {2025},
  publisher    = {National Consumer Law Center},
  address      = {Boston, MA},
  url          = {https://library.nclc.org/book/unfair-and-deceptive-acts-and-practices}
}

@article{kim2016data,
  title={Data-driven discrimination at work},
  author={Kim, Pauline T.},
  journal={William \& Mary Law Review},
  volume={58},
  pages={857},
  year={2016},
  publisher={HeinOnline}
}

@article{foggo2020algorithms,
  title={Algorithms, housing discrimination, and the new disparate impact rule},
  author={Foggo, Virginia and Villasenor, John},
  journal={Columbia Science \& Technology Law Review},
  volume={22},
  pages={1},
  year={2020},
  publisher={HeinOnline}
}

@inproceedings{verma2018fairness,
  title={Fairness definitions explained},
  author={Verma, Sahil and Rubin, Julia},
  booktitle={Proceedings of the international workshop on software fairness},
  pages={1--7},
  year={2018}
}

@article{caro2025differential,
  title={Differential Validity in Fair Lending},
  author={Caro, Spencer and Gillis, Talia B. and Nelson, Scott},
  journal={University of Chicago, Becker Friedman Institute for Economics Working Paper},
  number={2025-112},
  year={2025}
}

@article{schwemm2015fair,
  title={Fair Housing Litigation After Inclusive Communities: What's New and What's Not},
  author={Schwemm, Robert G.},
  journal={Columbia Law Review Sidebar},
  volume={115},
  pages={106},
  year={2015},
  publisher={HeinOnline}
}

@misc{ExecOrder2025Meritocracy,
  title        = {Restoring Equality of Opportunity and Meritocracy},
  howpublished = {\url{https://www.whitehouse.gov/presidential-actions/2025/04/restoring-equality-of-opportunity-and-meritocracy/}},
  note         = {Executive Order, April 23, 2025},
  author       = {{The White House}},
  year         = {2025},
  month        = {April}
}

@book{pridgen_consumer_law_2020,
  title        = {Consumer Law, Cases and Materials},
  edition      = {5th},
  author       = {Dee Pridgen, Jeff Sovern, Christopher L. Peterson},
  year         = {2020},
  publisher    = {West Academic Publishing},
  address      = {St. Paul, MN},
  isbn         = {9781642423099},
  pages        = {1205},
  url          = {https://faculty.westacademic.com/Book/Detail?id=298629}
}

@misc{nclc_maps_how_well_states_protect_consumers_2018,
  author       = {Carolyn Carter},
  title        = {{Consumer Protection in the States: A 50-State Evaluation of Unfair and Deceptive Practices Laws}},
  year         = {2018},
  month        = {March},
  organization = {National Consumer Law Center},
  url          = {https://www.nclc.org/resources/maps-how-well-do-states-protect-consumers/},
}

@article{gilles2023private,
  title={The Private Attorney General in a Time of Hyper-Polarized Politics},
  author={Gilles, Myriam},
  journal={Arizona Law Review},
  volume={65},
  pages={337},
  year={2023},
  publisher={HeinOnline}
}

@legal{usc_2000e2k,
  title     = {42 U.S.C. § 2000e-2(k)},
  jurisdiction = {United States},
  code      = {United States Code},
  year      = {2023},
}

@case{albemarle_paper_v_moody_1975,
  title       = {Albemarle Paper Co. v. Moody},
  court       = {U.S.\ Supreme Court},
  volume      = {422},
  reporter    = {U.S.},
  page        = {405},
  year        = {1975},
  docket      = {74-389},
}

@case{crypto_asset_v_hoard_2020,
  title       = {Crypto Asset Fund, LLC v. Hoard, Inc.},
  court       = {United States District Court for the Southern District of California},
  docket      = {No. 20-CV-438-MMA (AHG)},
  reporter    = {2020 WL 13556128},
  page        = {*1},
  year        = {2020},
  date        = {2020-06-19}
}

@misc{dfpi_gmo_global_desist_refrain_2022,
  title        = {Desist and Refrain Order: GMO Global, d/b/a Gmoglobal.com et al.},
  author       = {{California Department of Financial Protection and Innovation}},
  year         = {2022},
  month        = October,
  note         = {Order under California Financial Code § 90003 et seq.},
  url          = {https://dfpi.ca.gov/wp-content/uploads/sites/337/2022/10/D-R-GMO-Global.pdf}
}

@case{InclusiveCommunities2015,
  title       = {Texas Department of Housing and Community Affairs v. Inclusive Communities Project, Inc.},
  year        = {2015},
  court       = {Supreme Court of the United States},
  reporter    = {U.S.},
  volume      = {576},

}

@misc{FTCUnfairness1980,
  author       = {{Federal Trade Commission}},
  title        = {FTC Policy Statement on Unfairness},
  howpublished = {\url{https://www.ftc.gov/legal-library/browse/ftc-policy-statement-unfairness}},
  note         = {Dec. 17, 1980},
  institution  = {Fed. Trade Comm’n},
}

@misc{FTCA1994,
  title     = {Federal Trade Commission Act},
  author    = {Federal Trade Commission},
  year      = {1994}
}

@misc{FTCDeception1983,
  author       = {Federal Trade Commission},
  title        = {FTC Policy Statement on Deception},
  howpublished = {\url{https://www.ftc.gov/system/files/documents/public_statements/410531/831014deceptionstmt.pdf}},
  note         = {Oct. 14, 1983},
  institution  = {Fed. Trade Comm’n},
}

@misc{FTCPassport2022,
  title        = {Stipulated Order for Permanent Injunction, Monetary Relief, and Other Relief, Federal Trade Commission v. Passport Automotive Group},
  year         = {2022},
  howpublished = {No. 8:22-cv-02670-GLS (D. Md. Oct. 18, 2022)},
  institution  = {Fed. Trade Comm’n},
}

@article{rugaber_trump_consumer_protection_cease_2025,
  author       = {Christopher Rugaber},
  title        = {Trump administration orders consumer protection agency to stop work, closes building},
  journal      = {Associated Press},
  year         = {2025},
  month        = {Feb},
  day          = {9},
  url          = {https://apnews.com/article/trump-consumer-protection-cease-1b93c60a773b6b5ee629e769ae6850e9}
}

@misc{cfpb_strengthening_state_level_consumer_protections_2025,
  author       = {Consumer Financial Protection Bureau},
  title        = {Strengthening State-Level Consumer Protections},
  year         = {2025},
  month        = {Jan},
  url          = {https://files.consumerfinance.gov/f/documents/cfpb_strengthening-state-level-consumer-protections_2025-01.pdf}
}

@misc{nyag_james_sues_zelle_2025,
  author       = {{New York State Office of the Attorney General}},
  title        = {Attorney General James Sues Company Behind Zelle for Enabling Widespread Fraud},
  year         = {2025},
  month        = {Aug},
  day          = {13},
  note         = {Press release},
  url          = {https://ag.ny.gov/press-release/2025/attorney-general-james-sues-company-behind-zelle-enabling-widespread-fraud}
}

@article{sundheim_future_of_udaap_2025,
  author       = {Tony Sundheim \& Wolters Kluwer},
  title        = {The Future of UDAAP: Look to the States?},
  journal      = {Banking Exchange},
  year         = {2025},
  month        = {Mar},
  day          = {26},
  url          = {https://www.bankingexchange.com/news-feed/item/10278-the-future-of-udaap-look-to-the-states}
}

@legal{usc_5552a1,
  title        = {12 U.S.C. § 5552(a)(1)},
  code         = {United States Code},
  jurisdiction = {United States},
  year         = {2023},
  note         = {Dodd–Frank Act § 1042(a)(1)}
}

@article{cooper2016state,
  title={State Unfair and Deceptive Trade Practices Laws: An Economic and Empirical Analysis},
  author={Cooper, James C. and Shepherd, Joanna},
  journal={Antitrust Law Journal},
  volume={81},
  pages={947},
  year={2016},
  publisher={HeinOnline}
}

@misc{orrick_consumer_financial_services_2010,
  author       = {{Orrick, Herrington \& Sutcliffe LLP}},
  title        = {Consumer Financial Services, Volume 14, Issue 11 (Oct. 27, 2010)},
  year         = {2010},
  month        = {Oct},
  day          = {27},
  url          = {https://www.orrick.com/-/media/public/files/insights/2010/2010-10-27-cfs1411.pdf}
}

@article{pridgen2015wrecking,
  title={Wrecking Ball Disguised as Law Reform: ALEC's Model Act on Private Enforcement of Consumer Protection Statutes},
  author={Pridgen, Dee},
  journal={NYU Review of Law \& Social Change},
  volume={39},
  pages={279},
  year={2015},
  publisher={HeinOnline}
}

@article{faust2023regulating,
  title={Regulating Excessive Credit},
  author={Faust, Abigail},
  journal={Wisconsin Law Review},
  pages={753},
  year={2023},
  publisher={HeinOnline}
}

@article{cox2018strategies,
  title={Strategies of Public UDAP Enforcement},
  author={Cox, Prentiss and Widman, Amy and Totten, Mark},
  journal={Harvard Journal on Legislation},
  volume={55},
  pages={37},
  year={2018},
  publisher={HeinOnline}
}

@article{elengold2019consumer,
  title={Consumer Remedies for Civil Rights},
  author={Elengold, Kate Sablosky},
  journal={Boston University Law Review},
  volume={99},
  pages={587},
  year={2019},
  publisher={HeinOnline}
}

@misc{HUD2013DisparateImpact,
  title        = {Implementation of the Fair Housing Act’s Discriminatory Effects Standard},
  howpublished = {78 Fed. Reg. 11,460, 11,467 (Feb. 15, 2013) (codified at 24 C.F.R. pt. 100)},
  year         = {2013},
  institution  = {U.S. Department of Housing and Urban Development},
}

@misc{JointPolicy1994,
  title        = {Policy Statement on Discrimination in Lending},
  author       = {Department of Justice and Office of the Comptroller of the Currency and Board of Governors of the Federal Reserve System and Federal Deposit Insurance Corporation and Office of Thrift Supervision and National Credit Union Administration and Department of Housing and Urban Development},
  howpublished = {59 Fed. Reg. 18,266 (Apr. 15, 1994)},
  year         = {1994},
}

@case{Watson1988,
  title    = {Watson v. Fort Worth Bank \& Trust},
  reporter = {U.S.},
  volume   = {487},
  year     = {1988},
  court    = {Supreme Court of the United States},
}

@misc{EEOCUniformGuidelines1978,
  title        = {Uniform Guidelines on Employee Selection Procedures},
  author       = {{Equal Employment Opportunity Commission and Department of Labor and Department of Justice and Civil Service Commission}},
  howpublished = {43 Fed. Reg. 38,290 (Aug. 25, 1978) (codified at 29 C.F.R. pt. 1607)},
  year         = {1978},
}

@article{Tobia2022Disparate,
  author  = {Kevin Tobia},
  title   = {Disparate Statistics},
  journal = {Yale Law Journal},
  volume  = {132},
  year    = {2022},
  note    = {forthcoming, available at \url{https://papers.ssrn.com/sol3/papers.cfm?abstract_id=4079212}},
}

@case{WardsCove1989,
  title     = {Wards Cove Packing Co. v. Atonio},
  reporter  = {U.S.},
  volume    = {490},
  year      = {1989},
  court     = {Supreme Court of the United States},
}

@misc{RegB,
  title        = {Equal Credit Opportunity Act (Regulation B)},
  howpublished = {12 C.F.R. pt. 1002},
  year         = {2022},
  institution  = {Consumer Financial Protection Bureau},
  note         = {Implementing regulations for the Equal Credit Opportunity Act},
}

@case{FTC_FleetCor2022,
  title     = {Federal Trade Commission v. FleetCor Technologies, Inc.},
  reporter  = {F. Supp. 3d},
  volume    = {620},
  year      = {2022},
  court     = {N.D. Ga.},
  note      = {Summary judgment as to liability, noting that "the tiny, inscrutable print of the disclaimers does not cure the net impression."}
}

@misc{cfpb_statement_sj_res_57_2018,
  author       = {Consumer Financial Protection Bureau},
  title        = {Statement of the Bureau of Consumer Financial Protection on enactment of S.J. Res. 57},
  year         = {2018},
  month        = {May},
  day          = {21},
  url          = {https://www.consumerfinance.gov/about-us/newsroom/statement-bureau-consumer-financial-protection-enactment-sj-res-57/}
}

@misc{cfpb_targets_unfair_discrimination_2022,
  author       = {Consumer Financial Protection Bureau},
  title        = {CFPB Targets Unfair Discrimination in Consumer Finance},
  year         = {2022},
  month        = {Mar},
  day          = {16},
  url          = {https://www.consumerfinance.gov/about-us/newsroom/cfpb-targets-unfair-discrimination-in-consumer-finance/}
}

@misc{phillips_dissent_ftc_passport_oct2022,
  author       = {Noah Joshua Phillips},
  title        = {Dissenting Statement of Commissioner Noah Joshua Phillips Regarding FTC v. Passport Automotive Group, Inc. et al.},
  year         = {2022},
  month        = {Oct},
  day          = {14},
  url          = {https://www.ftc.gov/system/files/ftc_gov/pdf/Dissenting-Statement-of-Commissioner-Noah-Joshua-Phillips.pdf}
}

@legal{usc_12_5531c,
  title        = {12 U.S.C. § 5531(c)},
  note         = {Prohibiting unfair, deceptive, or abusive acts or practices},
  howpublished = {United States Code},
  year         = {2025},
  url          = {https://uscode.house.gov/view.xhtml?req=granuleid:USC-prelim-title12-section5531}
}

@legal{usc_15_45n,
  title        = {15 U.S.C. § 45(n)},
  note         = {Unfair methods of competition and unfair or deceptive acts or practices declared unlawful},
  howpublished = {United States Code},
  year         = {2025},
  url          = {https://uscode.house.gov/view.xhtml?req=granuleid:USC-prelim-title15-section45}
}

@misc{interagency_fair_lending_2009_Appendix,
  author       = {Office of the Comptroller of the Currency and Federal Deposit Insurance Corporation and Board of Governors of the Federal Reserve System and Office of Thrift Supervision and National Credit Union Administration},
  title        = {Interagency Fair Lending Examination Procedures --- Appendix},
  year         = {2009},
  url          = {https://www.ffiec.gov/pdf/fairappx.pdf},
  note         = {Available at \url {https://perma.cc/9U5K-QBV2}}
}

@misc{interagency_fair_lending_2009,
  author       = {Office of the Comptroller of the Currency and Federal Deposit Insurance Corporation and Board of Governors of the Federal Reserve System and Office of Thrift Supervision and National Credit Union Administration},
  title        = {Interagency Fair Lending Examination Procedures},
  year         = {2009},
  url          = {https://www.ffec.gov/PDF/fairlend.pdf},
  note         = {Available at \url {https://perma.cc/UH25-2LBC}}
}

@article{hirsch2014s,
  title={That's unfair-or is it: Big data, discrimination and the FTC's unfairness authority},
  author={Hirsch, Dennis D.},
  journal={Kentucky Law Journal},
  volume={103},
  pages={345},
  year={2014},
  publisher={HeinOnline}
}

@case{ChamberCFPB2023,
  title     = {Chamber of Commerce of the United States of America v. Consumer Financial Protection Bureau},
  court       = {U.S. District Court for the Eastern District of Texas},
  volume      = {691},
  reporter    = {F. Supp 3d},
  page        = {370},
  year        = {2023},
  docket      = {No. 6:22-cv-00381-JCB},

}

@legal{chamber_v_cfpb_2025,
  title        = {Order, Chamber of Commerce of the United States of America v. Consumer Financial Protection Bureau, No. 23-40650 (5th Cir. May 1, 2025)},
  court        = {United States Court of Appeals for the Fifth Circuit},
  docket       = {No. 23-40650},
  howpublished = {Order dismissing appeal pursuant to Fed. R. App. P. 42(b)},
  year         = {2025},
  date         = {2025-05-01},
  url          = {https://www.bloomberglaw.com/product/blaw/document/X1OER8N8QVR9ELQ9BLPV2958JHR}
}

@article{munnell1996mortgage,
  author       = {Alicia H. Munnell and Geoffrey M. B. Tootell and Lynn E. Browne and James McEneaney},
  title        = {Mortgage Lending in Boston: Interpreting HMDA Data},
  journal      = {American Economic Review},
  volume       = {86},
  number       = {1},
  pages        = {25--53},
  year         = {1996}
}

@case{neovi2010,
  title       = {Federal Trade Commission v. Neovi, Inc.},
  reporter    = {604 F.3d 1150},
  court       = {United States Court of Appeals for the Ninth Circuit},
  year        = {2010},
  pages       = {1150--1159},
}

@case{orkin1988,
  title       = {Orkin Exterminating Co., Inc. v. Federal Trade Commission},
  reporter    = {849 F.2d 1354},
  court       = {United States Court of Appeals for the Eleventh Circuit},
  year        = {1988},
  pages       = {1354--1360},
}

\newpage
\appendix

\renewcommand{\thefigure}{A.\arabic{figure}}
\setcounter{figure}{0}

\section{Appendix}
\label{sec:appendix}

\paragraph{More information on training set-up}

We train the following 20 configuration-based models:
\begin{itemize} 
    \item Logistic Regression: with all available features, numeric-only features, no credit-related features, and no job/employment related features.
    \item Random Forest: with 100 trees, and maximum depths of 3 or 5; tested on full and limited feature sets.
    \item Decision Tree: depth fixed at 5 across variations removing job features, credit features, or using limited features; also includes a shallow tree with depth 3.
    \item SVM: RBF and linear kernels, trained on both full and limited features.
    \item XGBoost: tested with tree depths of 3 and 5, as well as no-credit and limited-feature variants.
\end{itemize}

The remaining models are trained using grid search across a fixed hyperparameter space. The limited feature set contains only loan related features (e.g., loan amount, purpose, interest rate, etc.), excluding demographic, employment, and geographic information.

For example, for Random Forest models, we train models with with 100 trees, and maximum depths of 3 or 5; tested on full and limited feature sets, and for SVM models, we train both RBF and linear kernels, trained on both full and limited features.
The limited feature set contains only loan related features (e.g., loan amount, purpose, interest rate, etc.), excluding demographic, employment, and geographic information.

\autoref{fig:adult_fairlearn}, \autoref{fig:german_fairlearn}, and \autoref{fig:HMDA_fairlearn} 
show the same plots shown in \autoref{sec:tehcnical}, but with distinction as to which models were trained using Fairlearn \citep{bird2020fairlearn}.

\begin{figure}
    \centering
    \includegraphics[width=0.75\textwidth]{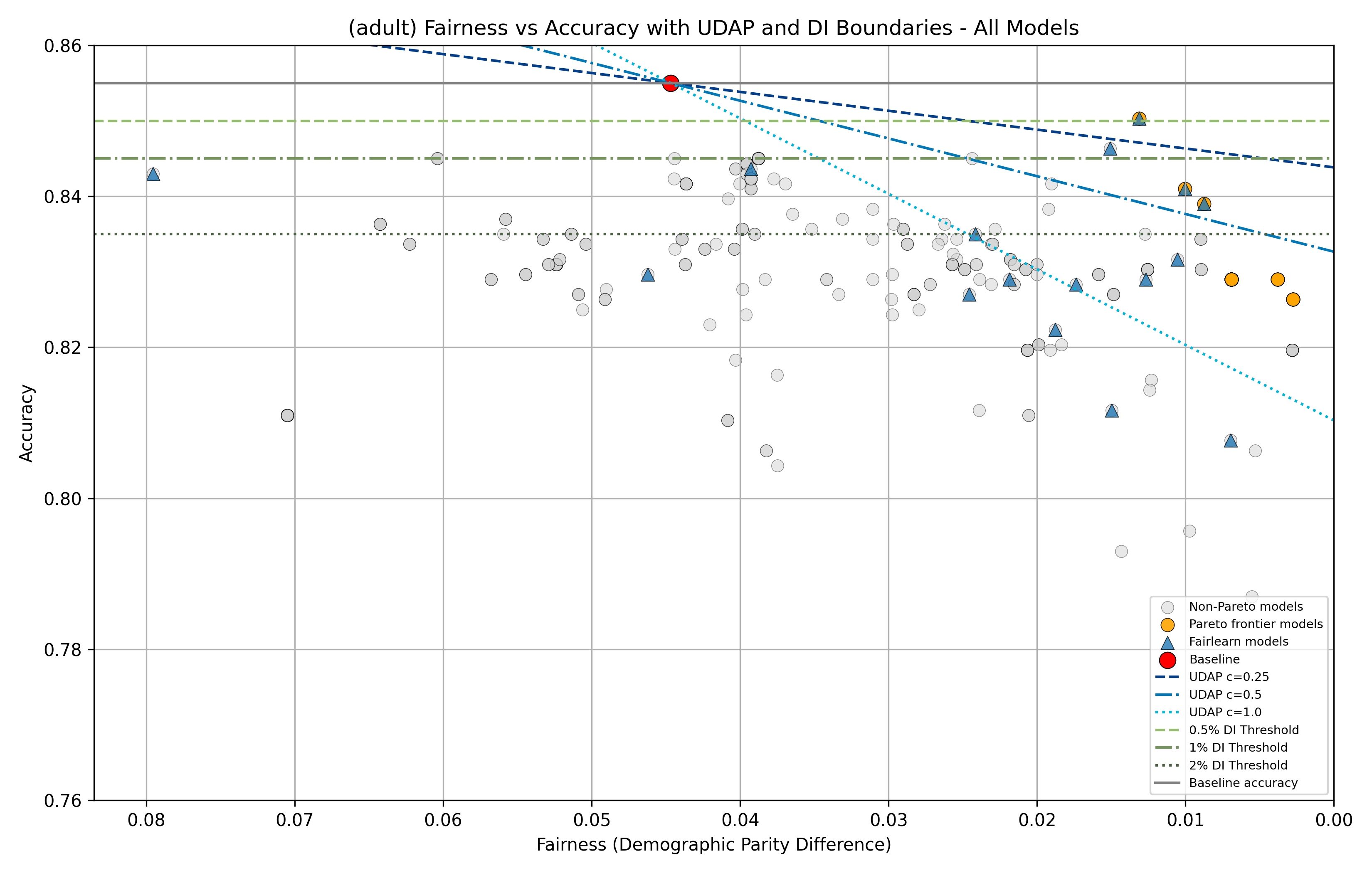}
    \caption{UDAP vs DI Analysis for finding alternatives on Adult Data, with Fairlearn indication. As before, X axis is demographic disparity \emph{decreasing} in severity from left to right, and on the y axis we have increasing accuracy. Models on the Pareto frontier are still in yellow, and the baseline model is denoted in red, but is also on the Pareto frontier. 
    Along with noting varying possible cut-offs for considering alternatives under DI and UDAP doctrines in blue and green, we denote models trained with Fairlearn with blue triangles.}
    \label{fig:adult_fairlearn}
\end{figure}

\begin{figure}
    \centering
    \includegraphics[width=0.75\textwidth]{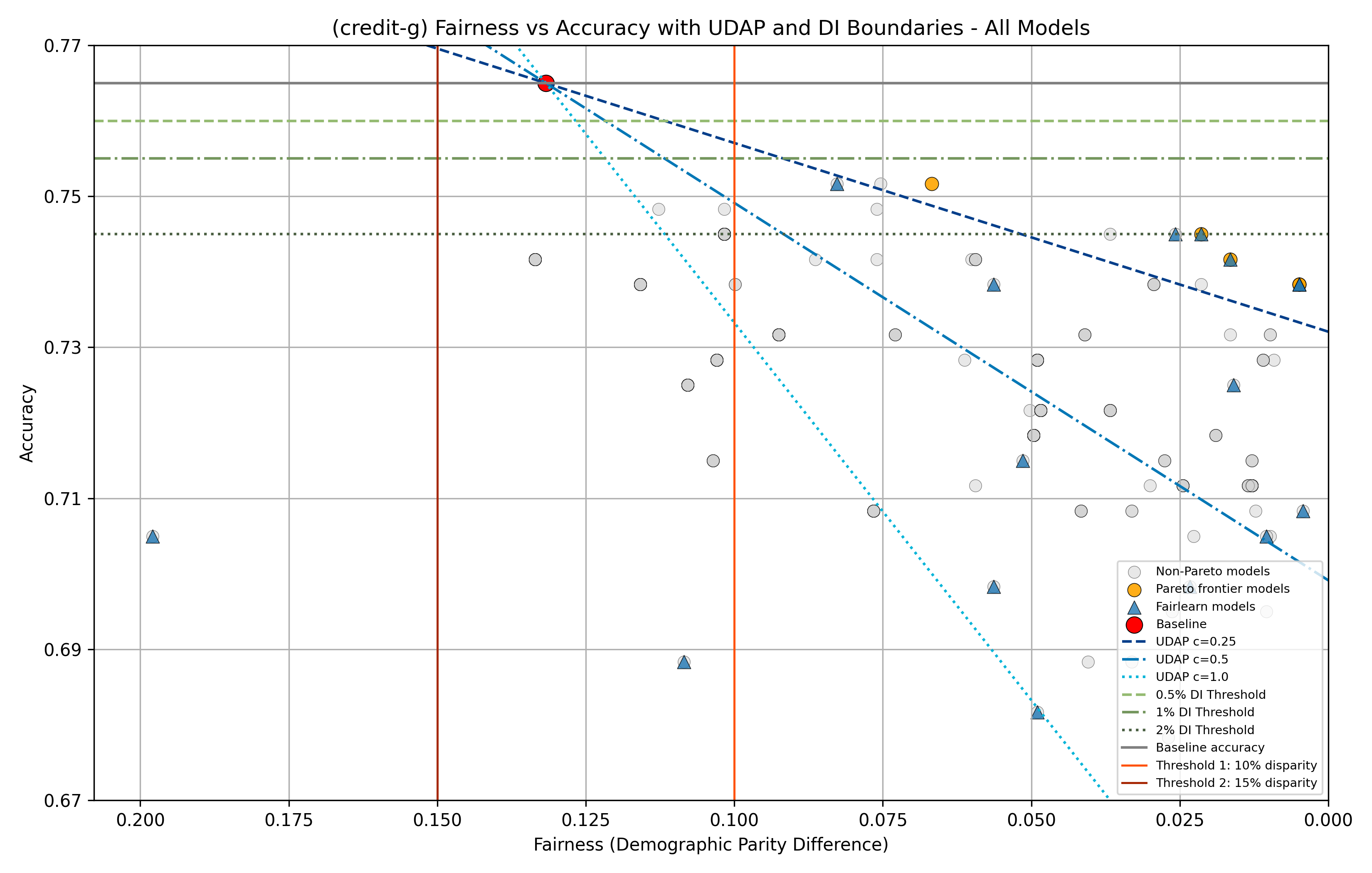}
    \caption{UDAP vs DI Analysis for finding alternatives on German Credit Data, with Fairlearn indication. As before, X axis is demographic disparity \emph{decreasing} in severity from left to right, and on the y axis we have increasing accuracy. Models on the Pareto frontier are still in yellow, and the baseline model is denoted in red, but is also on the Pareto frontier. 
    Along with noting varying possible cut-offs for considering alternatives under DI and UDAP doctrines in blue and green, we denote models trained with Fairlearn with blue triangles.}
    \label{fig:german_fairlearn}
\end{figure}

\begin{figure}
    \centering
    \includegraphics[width=0.75\textwidth]{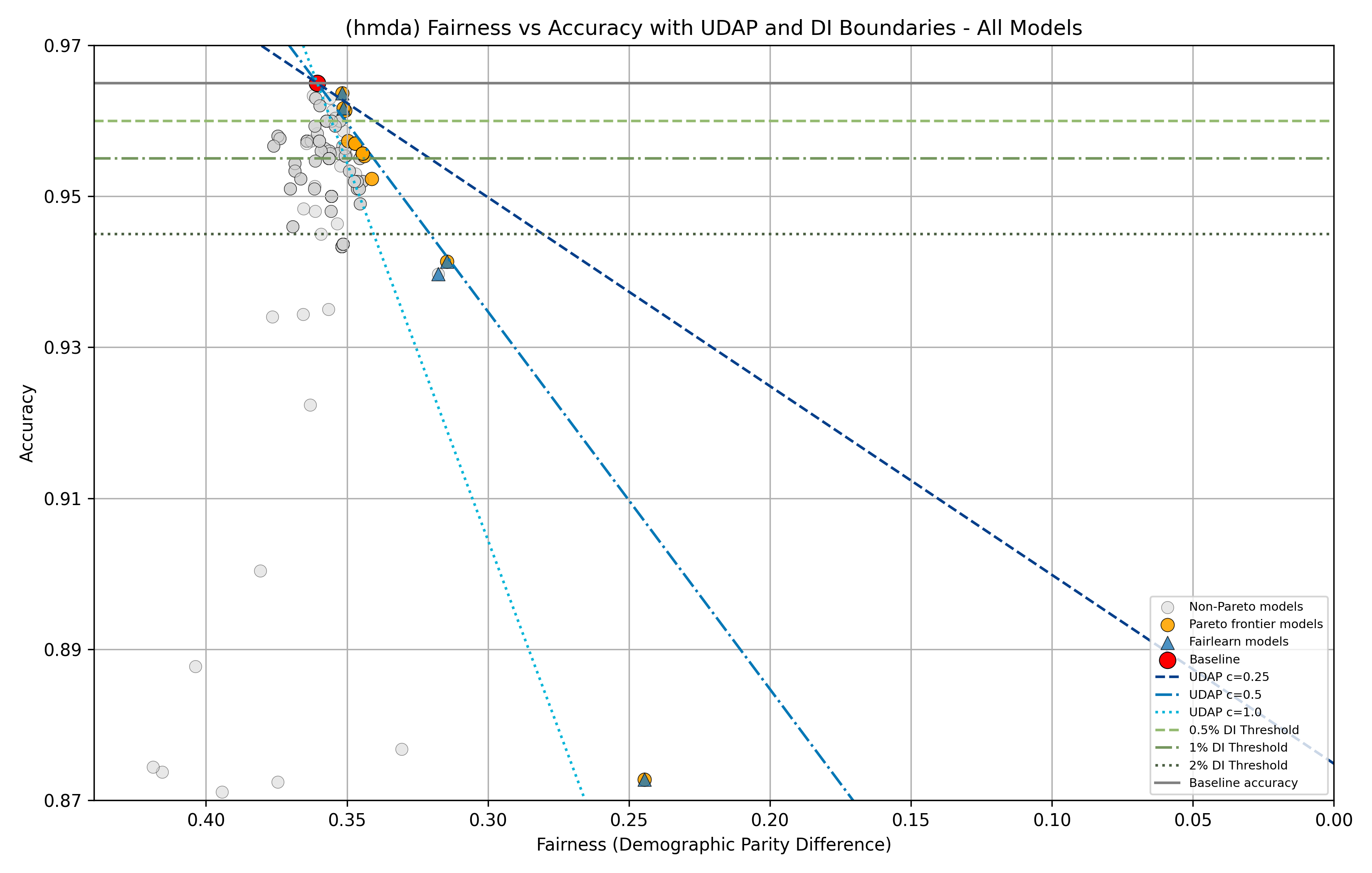}
    \caption{UDAP vs DI Analysis for finding alternatives on HMDA Data, with Fairlearn indication. As before, X axis is demographic disparity \emph{decreasing} in severity from left to right, and on the y axis we have increasing accuracy. Models on the Pareto frontier are still in yellow, and the baseline model is denoted in red, but is also on the Pareto frontier. 
    Along with noting varying possible cut-offs for considering alternatives under DI and UDAP doctrines in blue and green, we denote models trained with Fairlearn with blue triangles.}
    \label{fig:HMDA_fairlearn}
\end{figure}

\end{document}